\documentclass[12pt,preprint]{aastex}

\shorttitle{Magnetic Braking \& Protostellar Disks}
\shortauthors{Mellon \& Li}

\begin{document}

\title{Magnetic Braking and Protostellar Disk Formation: \\  
The Ideal MHD Limit}

\author{Richard R. Mellon$\!$\altaffilmark{1} \&  
Zhi-Yun Li$\!$\altaffilmark{1} 
\altaffiltext{1}{Astronomy Department, University of Virginia, 
Charlottesville, VA 22904; rrm8p, zl4h@virginia.edu} }

\begin{abstract}

Magnetic fields are usually considered dynamically important in star formation 
when the dimensionless mass-to-flux ratio is close to, or less than, unity
($\lambda~\lesssim~1$). We show that, in disk formation, the requirement is
far less stringent. This conclusion is drawn from a set of 2D
(axisymmetric) simulations of the collapse of rotating, singular isothermal 
cores magnetized to different degrees. We find that a weak field 
corresponding to 
$\lambda~ \sim 100$ can begin to disrupt the rotationally supported disk 
through magnetic braking, by creating regions of rapid, supersonic collapse 
in the disk. These regions are separated by one or more centrifugal barriers, 
where the rapid infall is temporarily halted. The number of centrifugal 
barriers increases with the mass-to-flux ratio $\lambda$. When $\lambda~ 
\gtrsim ~100$, they merge together to form a more or less 
contiguous, rotationally supported disk. Even though the magnetic
field in such a case is extremely weak on the scale of dense cores, 
it is amplified by collapse and differential rotation, to the 
extent that its pressure dominates the thermal pressure in both the 
disk and its surrounding region. For relatively strongly magnetized 
cores with $\lambda ~\lesssim ~10$, the disk formation is suppressed 
completely, as found previously. A new feature is that the mass 
accretion is highly episodic, due to reconnection of the magnetic 
field lines accumulated near the center. For rotationally
supported disks to appear during the protostellar mass accretion 
phase of star formation in dense cores with realistic 
field strengths, the powerful magnetic brake must be weakened, perhaps 
through nonideal MHD effects. Another possibility is to remove, 
through protostellar winds, the material that acts to brake the 
disk rotation. We discuss the 
possibility of observing a generic product of the magnetic braking, 
an extended circumstellar region that is supported by a combination 
of toroidal magnetic field and rotation --- a ``magnetogyrosphere'' --- 
interferometrically.

\end{abstract}

\keywords{accretion disks --- ISM: molecular clouds and magnetic 
fields --- MHD --- stars: formation}

\section{Introduction}

Disks are commonly observed around low-mass young stellar objects. They 
play a central role in star formation. It is generally 
believed that most of the mass of a Sun-like star is assembled through 
a disk \citep*{1987ARA&A..25...23S}. The disks may also be responsible for 
regulating the rotation rates of young stars, through disk-stellar 
magnetosphere interaction \citep{1991ApJ...370L..39K}, and 
for launching powerful 
jets and winds through the magnetocentrifugal mechanism 
\citep{2000prpl.conf..759K,2000prpl.conf..789S}. 
They are also the birthplace of planets.  

Low-mass stars form in dense cores of molecular clouds. The cores are
observed to rotate slowly \citep{1993ApJ...406..528G}. It is generally 
expected  
that the conservation of angular momentum during the core collapse 
would automatically lead to the formation of a rotationally supported 
disk. The expectation is borne out by both semi-analytic calculations 
\citep*[e.g.,][]{1984ApJ...286..529T} and numerical simulations (\citealp[e.g.,][]{1990ApJ...355..651B}; 
for reviews of early work, see \citealp{1995ARA&A..33..199B} and
\citealp{1998ASPC..148..314B}). 
However, the median value of the angular momenta 
measured for the cores \citep{1993ApJ...406..528G} is much higher than 
that inferred for circumstellar disks \citep{1993prpl.conf..521B}, 
by an order of magnitude or more \citep{1995ARA&A..33..199B}. Apparently, a 
considerable amount of angular momentum must be lost from the system 
in the process of core collapse and disk formation. 

A well-known mechanism for angular momentum removal is magnetic braking. 
Magnetic fields are believed to play an 
important role in the formation of low-mass stars in relative isolation  
\citep{1987ARA&A..25...23S,1993prpl.conf..327M,1999osps.conf..305M}.  
Polarization maps of 
dust continuum emission have revealed ordered magnetic fields in many
dense cores (see \citealp*{2006Sci...313..812G} for a spectacular example). The 
field strengths have been determined through Zeeman measurements in a 
number of them \citep{2007IAUS..237..141C}. The importance of magnetic 
field is usually measured by 
the mass-to-flux ratio $\lambda$ in units of the critical value 
$(2\pi G^{1/2})^{-1}$. For L1544,  
arguably the best studied starless core 
\citep{1998ApJ...504..900T,2007A&A...470..221C}, \citet{2000ApJ...537L.139C}
 inferred
$\lambda \sim 8$ using the line-of-sight components of the field
strength and column density. Correcting for projection effects may 
reduce $\lambda$ by a factor of 2 or more. In any case, there is 
ample observational evidence for both magnetic fields and rotation 
on the core scale. How the fields affect disk formation through 
magnetic braking of rotation is the focus of our investigation. 

Previous studies of magnetic braking have mostly concentrated on the 
core formation phase of cloud evolution prior to the formation of 
a central object \citep{1979ApJ...230..204M,1979MNRAS.187..337M,1989MNRAS.241..495N,1994ApJ...432..720B} 
or stopped shortly 
after a ``seed'' star has formed
(\citealp*{1998ApJ...502L.163T,2005A&A...435..385Z,2006ApJ...641..949B, 
2006A&A...457..371F,2007ApJ_MIM_sub};  
see, however, recent work by \citealp{2007A&A_HF_sub} and \citealp{2007A&A_HT_sub}).
What happens in the later, main protostellar mass accretion phase is 
less explored. \citet*[hereafter
  ALS03]{2003ApJ...599..363A} 
followed numerically the protostellar collapse of 
moderately strongly magnetized cores 
(with the dimensionless mass-to-flux ratio $\lambda \leq 10$), and 
found that rotationally supported disks were not able to form in the 
ideal MHD limit. In hindsight, the 
suppression of disk formation is not too surprising, given that any 
disk that did form would be magnetically linked to a slowly rotating 
envelope of much larger moment of inertia and would lose its angular
momentum quickly through magnetic braking. The braking is greatly 
enhanced by the equatorial pinching of field lines, which lengthens 
the magnetic lever arms \citep[ALS03,][]{2006ApJ...647..374G}. 

The catastrophic magnetic braking stood the traditional ``angular 
momentum problem'' of star formation on its head; the braking can 
be so efficient as to inhibit disk formation for a moderate level 
of cloud magnetization in the ideal MHD limit. The question is: 
under what condition will the  
rotationally supported disk reappear? To address this question, 
we go beyond ALS03 in two ways. First, we follow the collapse of 
magnetized, rotating dense cores using the spherical (as opposed 
to the cylindrical) version of the Zeus2D MHD code \citep{1992ApJS...80..753S,1992ApJS...80..791S}. 
The spherical geometry enables us to better resolve the 
angular distributions of the matter and magnetic field at small 
radii, where most of the magnetic braking is expected to occur. 
Second, we explore a wider range in the degree of cloud magnetization, 
$4 \leq \lambda \leq 400$. We find that the rotationally supported 
disk begins to be disrupted by magnetic braking for a dimensionless  
mass-to-flux ratio $\lambda\sim 100$, corresponding to an extremely 
weak field strength of order 1~$\mu$G for typical parameters of 
low-mass cores. An implication is that $\lambda=1$ is not the only
criterion for judging the dynamical importance of a magnetic field,
as is often assumed. A relatively weak field of $1 \ll \lambda~ 
\lesssim~ 100$ can dominate the angular momentum evolution in the main
accretion phase, because the field is amplified both by collapse 
and by rotation, and because the disk is braked over many rotation 
periods. Indeed, we find that even in the weakest field 
case that we have considered ($\lambda=400$), the field is 
amplified enough for the magnetic pressure to dominate the thermal 
pressure in the disk and its surrounding region, at least in 2D.  
The ideal MHD calculations provide a limiting case for 
calibrating calculations that include nonideal effects, 
particularly ambipolar diffusion, that are currently underway
(Mellon \& Li, in preparation).

The rest of the paper is organized as follows. We describe the problem 
setup and the code used for computation in \S~\ref{setup}. The numerical 
results are presented and discussed in \S~\ref{standard} through
\S~\ref{rotation}. In \S~\ref{discussion}, we discuss the generic
outcomes of the interplay between magnetic field and rotation during 
the protostellar mass accretion phase, including an extended region 
supported by a combination of toroidal magnetic field and rotation --- 
a ``magnetogyrosphere.'' Plausible 
observational evidence for such a structure in the Class 0 source 
IRAM 04191 is discussed. The last 
section \S~\ref{conclusion} contains the main conclusions. 

\section{Model Formulation}
\label{setup}

\subsection{Initial Condition}

The early stage of low-mass star formation can be divided 
conceptually into two phases: a prestellar and a protostellar 
phase. The two are separated by a pivotal state \citep{1996ApJ...472..211L}, 
in which the central density formally goes to infinity. 
The pivotal state marks the end of core formation and 
the beginning of protostellar mass accretion. It is 
characterized by a large (formally infinite) density contrast 
between the center and the edge of the core. Idealizing the pivotal (t=0) state 
as a static singular isothermal sphere (SIS), \citet{1977ApJ...214..488S} was 
able to obtain the well-known self-similar inside-out collapse 
solution for the protostellar phase of star formation. The 
present investigation is an extension of this work to those 
pivotal states that are both magnetized and rotating. 

The self-similarity of the SIS is preserved in the presence of 
a dynamically important magnetic field, provided that the 
mass-to-flux ratio along each field line is constant. This 
condition, termed ``isopedic''  by \citet{1996ApJ...472..211L}, 
is roughly satisfied in ambipolar diffusion-initiated 
formation of dense cores out of strongly magnetized clouds 
\citep{1989ApJ...342..834L,1993ApJ...415..680F}. In more weakly 
magnetized clouds, the nearly scale-free prestellar contraction 
of the isothermal gas right before the point mass formation 
would also drag the magnetic field of the dense core into 
a nearly isopedic configuration. ALS03 
found a family of self-similar solutions for the isopedically 
magnetized, rotating, singular isothermal pivotal state, which 
turned out to be toroids. They have carried out an initial 
study of the collapse of the pivotal
cores described by such solutions. Our investigation is a 
refinement and extension of theirs.

The singular isothermal toroids (SITs) are described, in a spherical 
polar coordinate system  ($r$, $\theta$, $\varphi$), by 
\begin{equation}
\rho(r,\theta)={a^2\over 2\pi G r^2}R(\theta),
\label{density}
\end{equation}
\begin{equation}
\Phi(r,\theta)={4\pi a^2 r\over G^{1/2}}\phi(\theta), 
\label{magflux}
\end{equation}
\begin{equation}
V_\varphi(r,\theta)=a~ v(\theta),
\label{rotationspeed}
\end{equation}
where $a$ is the isothermal sound speed, which we take to be a 
constant, and the functions $R(\theta)$, $\phi(\theta)$ and 
$v(\theta)$ describe the angular distributions of the density
$\rho$, magnetic flux $\Phi$, and rotation speed $V_\varphi$. 
We choose a fiducial value for the sound speed $a=0.3$~km/s  
(somewhat higher than $0.19$~km/s, the sound speed of molecular 
gas at 10~K), to account for the (typically subsonic) nonthermal
motions in dense cores \citep{1995mcsf.conf...47M}. The choice of $v(\theta)$ is 
discussed in depth in ALS03. Following their example, we set the 
rotation speed to a constant, i.e., $v(\theta)=v_0$. The density 
and magnetic flux distributions $R(\theta)$ and $\phi(\theta)$ 
are determined from a set of coupled ordinary differential 
equations (eqs.~[14] through [17] of ALS03) and associated 
boundary conditions. 

The toroid solutions are characterized by two free parameters,
$v_0$ and $H_0$. The parameter $v_0$ specifies how fast the
toroid rotates, while $H_0$ measures the amount of extra mass 
over the SIS that is supported by magnetic fields and rotation. 
For a given $v_0$, $H_0$ controls the magnetic field strength. 
ALS03 considered a number of combinations of $v_0$ and $H_0$,
including $v_0=0, 0.125, 0.25$ and $0.5$ for $H_0=0.25$, and 
$H_0=0.125, 0.25$ and $0.5$ for $v_0=0.25$. They 
found no evidence for a rotationally supported disk in any 
of their simulations, which have dimensionless mass-to-flux 
ratios ranging from $\lambda=2.77$ to 10. The result indicates 
that magnetic 
braking is very efficient, even for a magnetic field that is 
too weak to provide the bulk of support against self-gravity 
as a whole.  
On the other hand, a rotationally supported structure (a disk 
or a ring) is expected in the limit of zero magnetic 
field. As the field strength increases from zero, there must 
be a transition from a regime where a rotationally supported
structure is formed to a regime where it is completely 
suppressed. Quantifying how this transition occurs is one 
of the main goals of the current investigation. 

We will first focus on a specific combination of parameters, $v_0=0.5$ 
and $H_0=0.4$. For a fiducial isothermal sound speed $a=0.3$~km/s, 
the choice $v_0=0.5$ corresponds to an angular speed   
3~km~s$^{-1}$~pc$^{-1}$ on the scale of 
$0.05$~pc, the typical radius of a dense core \citep{1995mcsf.conf...47M}. The 
angular speed is within, although on the high side of, the range of velocity
gradient 
inferred observationally by \citet{1993ApJ...406..528G} for a collection 
of NH$_3$ cores (see their Fig.~1b). The relatively high rotation 
rate is chosen so that any rotationally support structure that 
may form is adequately resolved; slower rotations are expected to
be braked more easily. For $v_0=0.5$, the 
choice $H_0=0.4$ yields a dimensionless mass-to-flux ratio 
along each field line $\lambda=4$, consistent with the 
value inferred in the well-studied core L1544 \citep{2000ApJ...537L.139C}, 
after correcting for projection effects. 
It produces a distribution of (vertical) magnetic field on 
the equator 
\begin{equation}
B_{\rm eq}=24.5 \left({a\over 0.3~{\rm km/s}}\right)^2 
\left({0.05~{\rm pc} \over r}\right) (\mu G).
\label{bequator}
\end{equation} 

\begin{figure}
\epsscale{0.75}
\plotone{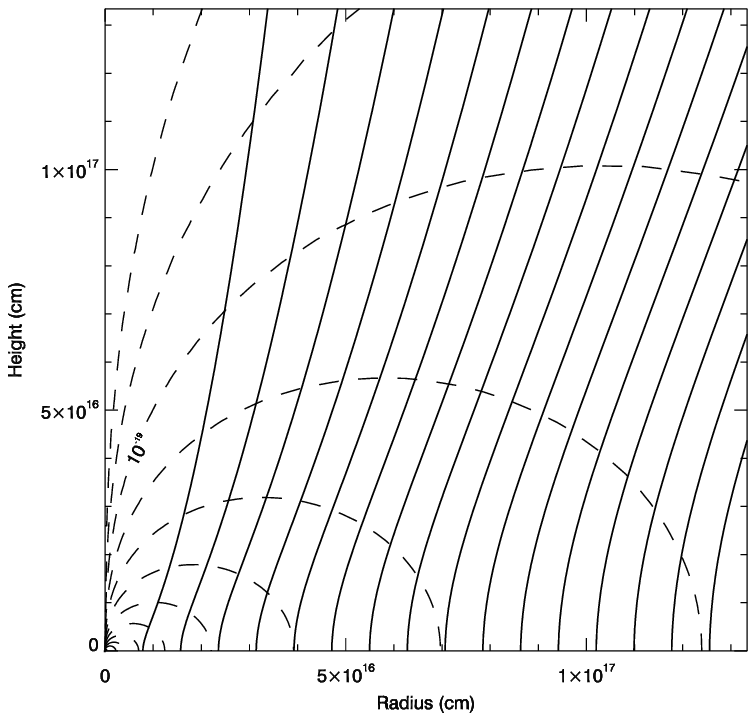}
\caption{Rotating, magnetized singular isothermal toroid (SIT) as the 
initial configuration for collapse calculation. Plotted are the 
isodensity contours (dashed lines, with two contours per decade 
and $10^{-19}$~g~cm$^{-3}$ labeled) and the lines 
of magnetic field (solid). 
} 
\label{initial}
\end{figure}

The rotating, magnetized singular isothermal toroid specified 
by $v_0=0.5$ and $H_0=0.4$ is shown in Fig.~\ref{initial}. Compared 
with the SIS of the same isothermal sound speed, the SIT has an 
enhanced 
density on the equator (by a factor of 2.25), due to extra support 
by the magnetic field and, to a lesser extent, rotation. The density 
is depleted in the polar region as matter settles along 
the field lines, producing the toroid appearance. The magnetic 
pressure dominates the 
thermal pressure in the evacuated polar region, within a half 
opening angle of $40^\circ$ of the axis. Outside the region (where 
most of matter resides), the thermal pressure dominates, with a 
plasma-$\beta=7.68$ on the equator. This magnetic field, although 
too weak to prevent the core from collapsing inside out, is strong 
enough to suppress the formation of a rotationally supported disk 
completely, as we demonstrate below (see also ALS03). Indeed, 
even a much weaker magnetic field can modify the process of disk 
formation significantly in the ideal MHD limit. 

To quantify the effects of field strength on magnetic braking and 
disk formation, we will reduce the field strength of the toroid 
shown in Fig.~\ref{initial} by various constant factors everywhere, keeping 
the distributions of density and rotation speed fixed. We consider 
a wide range in the degree of magnetization, corresponding to 
$\lambda=400$, 200, 133, 80, 40, 20, 13.3, 8, and 4 (see Table~1). For 
reference, $\lambda=100$ corresponds to a field strength of merely $1~\mu$G 
on the core scale of $0.05$~pc (for the fiducial value of sound 
speed), 
well below the median field strength ($\sim 6~\mu G$) in the 
cold neutral structures of HI gas \citep{2005ApJ...624..773H}. The 
field strength in dense cores of molecular clouds 
is unlikely to be as low as $1~\mu$G, but we include cases
with high values of $\lambda$ to illustrate the transition from 
disk formation to suppression. For more turbulent (perhaps
massive star formation) regions where the effective sound speed 
is much higher than 0.3~km/s, even the very large $\lambda$ 
cases may become relevant, as we emphasize in the discussion 
section. 

A potential drawback of reducing the field strength without changing
the distributions of density and rotation speed correspondingly 
at $t=0$ is that the initial configuration is out of force balance. 
The weakening of magnetic support will induce large-scale infall 
motions at times $t >0$. Infall motions on the core scale are 
observed, however, in a number of dense cores, including L1544 
\citep{1998ApJ...504..900T}, with speeds typically of order half the 
sound speed. Such a subsonic motion is present in our standard 
model to be discussed in depth in \S~3. The reduction of field 
strength by a uniform factor everywhere at $t=0$ has the added 
advantage that the core collapse at $t > 0$ remains self-similar, 
which provides a powerful check on the numerical solution. The
same advantage is preserved when we reduce the fiducial rotation 
speed at $t=0$ by a uniform factor everywhere. The effects of 
initial rotation speed are discussed separately from those of 
initial field strength in \S~\ref{rotation}. 

\begin{deluxetable}{llll}
\tablecolumns{4}
\tablecaption{Model Parameters\label{tab:model}}
\tablewidth{0pt}
\tablehead{
  \colhead{Model}     & \colhead{$\lambda$} 
& \colhead{$v_0$} & \colhead{$B_c^a~(\mu G)$} 
}
\startdata
B0   & $\infty$    & 0.5   & 0.0  \\
B1   & 400  & 0.5   & 0.25 \\
B2   & 200  & 0.5   & 0.49 \\
B3   & 133  & 0.5   & 0.74 \\
B4   & 80   & 0.5   & 1.23 \\
B5   & 40   & 0.5   & 2.45 \\
B6   & 20   & 0.5   & 4.90 \\
B7$^b$  & 13.3 & 0.5   & 7.35 \\
B8   & 8    & 0.5   & 12.3 \\
B9   & 4    & 0.5   & 24.5 \\

R0   & 13.3 & 0.0   & 7.35 \\
R1   & 13.3 & 0.125 & 7.35 \\
R2   & 13.3 & 0.25  & 7.35 \\

\enddata
\tablecomments{(a) Initial equatorial field strength (see 
equation~[\ref{bequator}]). (b) Model B7 is the standard model to be
discussed in depth in \S~\ref{standard}.}
\end{deluxetable}

\subsection{Boundary Conditions}

The protostellar phase of low-mass star formation in magnetized cores
is challenging to simulate in the ideal MHD limit. Once a 
central point mass is formed, it will carry along with it a 
finite amount of magnetic flux. The trapped flux would formally produce 
a split magnetic monopole \citep{1993ApJ...417..243G,1997ApJ...475..237L}, 
causing the field strength to increase rapidly with decreasing 
radius ($\propto r^{-2}$). The rapid 
increase of field strength makes the Courant condition associated
with the Alfv{\'e}n speed prohibitively small close to the origin, 
where the grid is necessarily very fine, especially in the $\theta$
direction of the adopted spherical polar coordinate system. 

To alleviate the numerical problem associated with the formation 
of a split magnetic monopole, we put the inner boundary of our
computation domain at a fixed radius, $r_i=10^{14}$~cm, or 
$6.7$~AU. We are unable to follow the evolution of the flow inside 
$r_i$. However, for our adopted initial conditions, only 
a small fraction of the core mass has a low enough specific 
angular momentum to fall through the inner surface (or `` inner
hole'') without magnetic braking; for the majority of the 
mass, extensive braking
must occur in the computation domain to bring it to the inner 
surface. On this surface, we impose the standard ``outflow'' 
boundary conditions for the hydrodynamic quantities, i.e., the 
density, energy and tangential (to the surface) velocity 
components are copied from the first active zones into the 
ghost zones along the radial direction. The radial velocities 
in the ghost zones are set to the lesser of the radial 
velocity in the first active zone and zero. These 
conditions are a standard feature already implemented in the 
Zeus2D code \citep{1992ApJS...80..753S,1992ApJS...80..791S} that is employed for our 
simulations.

The inner boundary conditions for the magnetic field require
special attention. Since the poloidal components of the 
magnetic field are evolved through the method of constrained
transport \citep{1988ApJ...332..659E} in Zeus2D, their boundary 
conditions are imposed on the 
electromotive force (EMF) ${\bf \epsilon}={\bf V}\times 
{\bf B}$. We apply the standard Zeus2D ``outflow'' boundary 
conditions on the EMF to evolve the poloidal magnetic field. 
They produce a poloidal field that varies smoothly across 
the inner surface. The toroidal component of the magnetic 
field $B_\varphi$ is not evolved 
through the  method of constrained transport, and it is 
possible to impose boundary conditions on $B_\varphi$ 
directly. We pick the simplest possible condition 
for the toroidal field in the ghost zones: $B_\varphi =0$. 
It is in effect a torque-free boundary condition. The choice 
is motivated in part by the realization that, by the time a 
piece of core material reaches the (small) inner surface, most 
of its angular momentum would have been stripped away already. 
The remaining angular momentum would be too little to significantly 
twist the strong (split-monopole) field lines that thread the 
inner surface \citep[see][for a related discussion]{2006ApJ...647..374G}. 
Other prescriptions are possible. For example, \citet{2001MNRAS.322..461S} 
enforced the negative stress condition $B_r B_\varphi
\leq 0$, whereas \citet{2001ApJ...548..348H} set the transverse 
components of the magnetic field to zero in their simulations
of black hole accretion in a cylindrical coordinate system. 
We have experimented with other prescriptions for $B_\varphi$ 
in the ghost zones, including copying values from the first 
active zones, which produced a somewhat stronger outflow 
along the axis compared with the torque-free case. The 
outflow is weak by the standard of the powerful jets and 
winds observed in young stars (ALS03), and will be easily 
masked by them. In any case, the simple torque-free boundary 
condition does not 
directly affect the braking of the bulk of the core material 
that accretes through the equatorial region (including any 
disk that may form), the main focus of our investigation. 

Boundary conditions are also required at the outer boundary. We 
set the outer radius of the computation domain to $r_o=2\times 
10^{17}$~cm, slightly 
larger than the fiducial core radius of $0.05$~pc. Within this 
radius, there is $3.8$~$M_\odot$ of material for the initial 
density distribution shown in Fig.~\ref{initial} (for 
$a=0.3$~km/s), more than enough to form a typical low-mass 
star of 0.5~$M_\odot$. On the outer boundary, we impose the 
standard Zeus2D ``outflow'' boundary conditions for all 
quantities, including $B_\varphi$. 

We assume that the core collapse remains symmetric with respect 
to the polar axis and the equatorial plane at all times. We adopt 
the standard Zeus2D boundary conditions for the axis. To enforce 
the equatorial boundary conditions, we mirror all quantities
(including the three components of the magnetic field) in 
the upper quadrant to the lower quadrant at the beginning
of each time step, and evolve the governing equations in 
both quadrants. This implementation of equatorial boundary
conditions, although somewhat more expensive than the one
already in Zeus2D, guarantees that the symmetry is enforced
exactly. Spontaneous symmetry breaking is observed in MHD 
simulations of black hole accretion \citep{2001ApJ...548..348H,2001MNRAS.322..461S}. 
We will relax the equatorial symmetry 
in future 3D simulations. 

\subsection{Code Modification and Problem Setup} 

For our axisymmetric problem, we use the Zeus2D MHD code of 
\citet{1992ApJS...80..753S,1992ApJS...80..791S} 
in spherical polar coordinate system. The code is 
well suited for our investigation of magnetic braking, since the
propagation of (torsional) Alfv\'en waves is treated accurately 
with the method of characteristics. The main modification that we 
made to the code is to improve the method for obtaining the values
of gravitational potential at the inner and outer boundaries; they 
are needed for solving the Poisson equation for the self-gravity 
of the material inside the computational domain. The method of 
multipole expansion in Zeus2D performs poorly when mass is near a 
boundary and requires a large number of terms to accurately determine 
the gravitational potential.  Instead, we transform Green's function 
from the multipole solution to an elliptic integral solution 
\citep{1999ApJ...527...86C}. 
We can then directly sum up 
the contributions of the masses in all active cells to the 
gravitational potential at a given location on the boundary, as 
in \citet{2004ApJ...616..364F}. We also modified the equation of state 
to change smoothly from isothermal to adiabatic 
around a critical density of $10^{-13}$~gm~cm$^{-3}$, to mimic 
the effects of radiation trapping at high densities. 

To ensure good resolution on both large and small scales, a 
logarithmically spaced grid is used in the radial direction. 
We divide the computational domain between the inner 
($r_i=10^{14}$~cm) and outer radius ($r_o=2\times 10^{17}$~cm) 
into 120 shells, each having a width that is 6\% larger than 
that of the shell interior to it. Since 
both $\Delta r$ and $r\Delta \theta$ 
vary linearly with the radius, the grid does not contain cells 
of large aspect ratios often seen in logarithmic grids in 
cylindrical coordinate system. The width of the first shell 
is $1.10\times 10^{13}$~cm, much smaller than the inner radius
$r_i$. In the polar direction, we divide the region between 
the upper ($\theta=0$) and lower axis ($\theta=\pi$) into 119 
uniform wedges, each with an opening angle of $1.5^\circ$. 
As mentioned earlier, half of the grid (between $\theta=\pi/2$ 
and $\pi$) is used for enforcing mirror symmetry across the 
equator. 

We begin the collapse calculation with a static, dense core. 
To initiate the collapse, we add a point 
mass of $M_0=3 a^3 r_i/G$ (where $a$ is the isothermal
sound speed and $r_i$ the radius of the inner hole) at $t=0$. 
This mass is 1.5 times the mass enclosed within $r_i$ for 
a SIS of sound speed $a$, and slightly larger than that 
enclosed within the same radius for the magnetized SIT shown 
in Fig.~\ref{initial} ($2.8 a^3 r_i/G$). As the collapse 
proceeds, the density is depleted preferentially in the 
polar region, where the magnetic field is strong and matter 
slides more or less freely along the field lines to the center 
(ALS03). To prevent the Alfv\'en time step from dropping to
prohibitively small values, we adopt a density floor of $10^
{-19}$~gm~cm$^{-3}$ on the inner half of the (logarithmic) 
radial grid (within $5.3\times 10^{15}$~cm). The occasional 
use of density floor in the polar region is expected to 
have little dynamical effect, since the dynamics there are 
controlled by the magnetic field. We have experimented with 
different choices of the floor value and confirmed the
expectation. 

\section{Standard Model} 
\label{standard}

The rotating, magnetized core that we adopt as the initial configuration 
for the collapse calculation is specified by two parameters: the 
dimensionless mass-to-flux ratio $\lambda$ and the rotation speed 
$v_0$ in units of the sound speed. We will explore a wide range of 
values for $\lambda$ in \S~\ref{fieldstrength} and several values 
for $v_0$ in \S~\ref{rotation}. In this section, we focus on a model 
with a particular combination of parameters $\lambda=13.3$ and
$v_0=0.5$
(Model B7 in Table~1); 
it serves as the standard against which others models are compared. 
The choice $\lambda=13.3$ corresponds to a moderate field strength of 
$7.35$~$\mu$G on the scale of the typical core radius $0.05$~pc 
(not much higher than the median field strength of $\sim 6.0$~$\mu$G 
inferred by \citet{2005ApJ...624..773H} for the cold neutral 
HI structures), and $v_0=0.5$ is chosen to ensure that any rotationally 
supported structure that may form is large enough to be adequately 
resolved. More importantly, it turns out that the combination 
yields a collapse solution that is roughly self-similar in time, 
as expected from the self-similar initial configuration adopted. 
The self-similarity gives us confidence on the numerically obtained 
solution. 

\subsection{Collapse Solution at a Representative Time}

Fig.~\ref{snap} shows a snapshot of the standard model at a 
representative time $t=6.4\times 10^{11}$~sec (or $\sim 2\times 
10^4$~yrs after the initiation of collapse), comparable to the
typical ages estimated for Class 0 sources. The snapshots at 
other times are largely similar (to be quantified below), which 
motivates us to introduce a set of dimensionless variables:
\begin{equation}
\alpha =4\pi G t^2 \rho(r,\theta,t),\ 
{\bf b}={G^{1/2} t\over a} {\bf B}(r,\theta,t),\ 
{\bf v}={{\bf V}(r,\theta,t)\over a},\ x={r\over a t}.
\label{dimensionless}
\end{equation}
If the collapse is strictly self-similar, the dimensionless density 
$\alpha$, magnetic field ${\bf b}$ and velocity ${\bf v}$ would be 
functions of the dimensionless self-similar radius $x=r/(at)$ and 
polar angle $\theta$ only, and the solution would look identical at 
different times except for a scaling factor. Equation (5) can be 
used to obtain the dimensional units for any dimensionless quantity
to be used below, particularly in figures; the plotted quantities 
will be dimensionless unless noted otherwise explicitly. 

Broadly speaking, there are four dynamically distinct regions. At 
large (cylindrical) distances, the core material collapses inward, 
dragging along magnetic field lines, particularly in the equatorial 
region. We term this region the ``collapsing envelope'' or ``Region 
I.'' Most of the collapsing material from the envelope is channeled 
into the equatorial region, forming a flattened ``pseudodisk'' or 
``Region II.'' The pseudodisk is squeezed from above by an expanding 
``magnetic bubble'' (or ``Region III'') that lies at intermediate 
latitudes. The pseudodisk supports the bubble from below. On the 
other side of the bubble, closer to the polar axis, lies ``Region 
IV,'' where matter slides down well-ordered field lines, forming 
a low-density ``polar funnel.'' We consider the outflowing region 
near the axis part of ``Region III;'' it is a continuation of the 
bubble to large distances. 

The boundaries between the various regions show up most 
clearly in the map of the magnetic twist, defined as the ratio of 
the toroidal to poloidal field strength $B_\varphi/\vert B_p \vert$ 
(see Fig.~\ref{twist}). To be specific, we define the region 
dominated by the (negative) toroidal magnetic field (i.e., 
$-B_\varphi > \vert B_p \vert$) as the bubble (``Region 
III''). It is surrounded by the polar funnel to the left, the 
collapsing envelope to the right, and the pseudodisk from 
below. The collapsing envelope is separated from the bubble 
by a ``magnetic wall,'' produced by the deceleration of the 
magnetized collapsing envelope against the bubble; the  
deceleration leads to a pileup of the field lines. The 
decelerated material slides along the compressed field lines 
nearly vertically towards the equator. Before hitting 
the equator, the downward moving material is deflected 
towards the origin by a strong (thermal) pressure gradient 
to join the pseudodisk beneath the magnetic bubble. 

The equatorial part of the collapsing envelope joins onto the 
pseudodisk more directly. It slows down near the outer edge of
the pseudodisk before being reaccelerated towards the center 
by gravity. The deceleration and reacceleration show up more 
clearly in Fig.~\ref{vrvphi_eq}, where we plot the radial 
component of the velocity on the equator as a function of 
radius. The curve of radial speed has a peak around $x_{\rm s} 
\approx 0.5$, where $v_r$ is close to zero; indeed, the gas 
there is actually slowly expanding rather than collapsing. 
We identify the region near the stagnation radius $x_{\rm s}$ 
as a ``magnetic barrier'', as opposed to a ``centrifugal 
barrier.'' The reason is that, at this location, the rotation
speed is well below that needed for the centrifugal force to 
balance the gravity (see Fig.~\ref{vrvphi_eq}, where both speeds
are plotted).
There is tantalizing evidence for such a structure 
in the Class 0 source IRAM 04191, as we discuss in 
\S~\ref{magnetogyrosphere}. It bounds the pseudodisk from 
outside. Note that the infall speeds at large distances are 
close to half the sound speed, comparable to that inferred 
observationally in the dense core L1544 (as mentioned earlier)
and in the envelope of IRAM 04191 \citep{2002A&A...393..927B}, 
except near the outer boundary, where the edge effect becomes 
significant. The large-scale infall is induced by the reduction 
of the full field strength at $t=0$ (from $\lambda=4$ to 
13.3), which weakens the initial magnetic support relative 
to self-gravity. 

After passing through the stagnation region at the magnetic 
barrier, the pseudodisk material on the equator collapses quickly 
inward. The rapid infall leads to a spinup of the collapsing 
material, as shown in Fig.~\ref{vrvphi_eq}. 
The inward increase in rotation speed 
is slowed down around a dimensionless radius $x_{\rm c1}\sim 
0.1$, and levels off near $x_{\rm c2}\sim 0.05$. A glance at
Fig.~\ref{twist} shows that this is the region where the magnetic 
bubble is initiated. In this region, the radial 
infall is temporarily slowed down, as a result of increased 
centrifugal support; in other words, a centrifugal barrier 
is encountered. The slowdown of infall allows more time for 
the magnetic braking to operate. Interior to  $x_{\rm c2}$ is 
a strongly magnetized region dominated by the central split 
monopole. Here, the rotation speed drops steadily with decreasing 
radius, as a result of efficient braking by the strong magnetic 
field. The loss of centrifugal support causes the equatorial 
material to resume its rapid collapse towards the center. The 
rapid infall and slow rotation leave no doubt that a rotationally
supported disk is not produced in this standard model. In particular, 
the rotation speed at small radii is well below that needed to
provide centrifugal support against gravity (compare the top
two curves in Fig.~\ref{vrvphi_eq}). Upon arriving at the inner 
hole, the infall material has lost most of its original angular 
momentum along the way. The questions is: how does the stripping 
of angular momentum occur mechanically? 

The key to answering the above question lies in the magnetic 
bubble (Region III). It is powered by the pseudodisk (Region 
II). The material in the pseudodisk spins up as it collapses, 
producing a growing toroidal magnetic field, which in turn 
leads to a buildup of magnetic pressure. The 
over-pressure in the region is relieved by expansion 
along the path of least resistance. Expansion is restricted to 
the left of the region or vertically up by the strong monopole 
field. It cannot freely expand horizontally outward either; 
it is contained by the ram pressure of the infalling material. 
These constraints force the over-pressured region to expand 
outward at an acute angle with respect to the equatorial 
plane, producing a bubble where the bulk of the angular 
momentum stripped from the collapsing pseudodisk is stored. 

\subsection{Approximate Self-Similarity in Time Evolution}
\label{ss}

Self-similarity not only provides a check on the numerically obtained 
solution, but also enables us to discuss the solution in a
time-independent way. A good indicator of the self-similarity of 
the collapse solution is the evolution of the central point mass
$M_0$. In a strictly self-similar solution 
(such as Shu's inside-out collapse solution), 
the mass should increase linearly with time. The 
change of $M_0$ with time for the standard model is shown in 
Fig.~\ref{pointmass}. Overall, there is a nearly linear increase 
of the point mass with time, except near the origin. The initial 
deviation is caused by the finite size of the inner hole (which 
breaks the self-similarity of the initial configuration), and 
the point mass that was put at the center at $t=0$ to induce 
collapse. After a short period of initial adjustment (of 
order $10^{11}$~sec), the mass-time curve becomes 
remarkably straight, although there are low-amplitude oscillations. 
As we discuss in detail in \S~\ref{strongfield}, the oscillation
is caused by magnetic flux accumulation near the center, which 
produces a split magnetic monopole that reconnects episodically 
across the equatorial plane. The reconnection events make it 
impossible for the collapse to settle into a strictly steady 
state in the self-similar space. Their influence on the collapse
solution increases with the strength of the magnetic field. 
For the standard model, the field is still weak enough that the 
self-similarity is preserved approximately.

To demonstrate the approximate self-similarity more vigorously,  
we use a closure relation that any self-similar collapse 
solution must satisfy. The relation is derived from the equation 
of mass conservation
\begin{equation}
{\partial M(r,t)\over \partial t}=-2\pi \int^\pi_0 \rho r^2 
\sin(\theta)V_r d\theta,
\label{masscons}
\end{equation}
where $M(r,t)$ is the mass enclosed within a radius of $r$ at the instant 
of time $t$. It is given by
\begin{equation}
M(r,t)=M_0(t)+2\pi \int_0^r \int^\pi_0 \rho {\tilde r}^2 
\sin(\theta) d\theta d {\tilde r},
\label{mass}
\end{equation}
where $M_0$ is the point mass at the center. In terms of the dimensionless 
self-similar variables introduced in equation~(\ref{dimensionless}), the 
mass $M(r,t)$ can be written into
\begin{equation}
M(r,t)={a^3 t\over 2 G} m(x), 
\label{mass_scaling}
\end{equation}
where
\begin{equation}
m(x)=m_0+\int_0^x \int_0^\pi \alpha({\tilde x},\theta) {\tilde x}^2 
\sin(\theta) d\theta d{\tilde x},
\label{mass_dimensionless}
\end{equation}
is the dimensionless mass enclosed within a dimensionless radius $x$,
and $m_0$ is the dimensionless point mass at the center. Using 
equation~(\ref{mass_scaling}), we can express the left-hand side of 
equation~(\ref{masscons}) as 
\begin{equation}
{\partial M(r,t)\over \partial t} = {a^3\over 2G} \left[m(x)-x 
m^\prime(x) \right],
\label{chainrule}
\end{equation}
where $m^\prime=dm/dx= \int_0^\pi \alpha(x,\theta) x^2 \sin(\theta)
d\theta$ from equation~(\ref{mass_dimensionless}). Equating the 
right-hand side of equation~(\ref{chainrule}) with the right-hand 
side of equation~(\ref{masscons}) yields
\begin{equation}
m(x) = \int_0^\pi \alpha(x,\theta) x^3 \sin(\theta) d\theta+
\int_0^\pi \alpha(x,\theta) x^2 \sin(\theta) (-v_r) d\theta
=\int_0^\pi \alpha(x,\theta) x^2 \sin(\theta) (x-v_r) d\theta,
\label{masscons_dimensionless}
\end{equation}
which is the equation of mass conservation in the self-similar space.  
It has a simple physical meaning. The term on the left-hand side is 
the dimensionless mass enclosed within a sphere of radius $r=x a t$ 
that expands linearly in time (for a fixed $x$). This mass can be 
increased due to either 
the expanding volume (which is proportional to the first term on the 
right-hand side of the first equality) or advection of material 
into the sphere from outside 
(proportional to the second term).  Note that the combination $x-v_r$
on the right hand side of the second equality is simply the speed with 
which the sphere is expanding relative to 
the material that moves at a radial speed of $v_r$. 

The dimensionless mass distribution $m(x)$ can be evaluated in another
way, using equations~(\ref{mass_dimensionless}), which states that the 
mass inside a given radius is simply the sum of all mass (including 
the point mass) up to that 
radius. In a strictly self-similar solution, the masses determined 
independently from equations~(\ref{mass_dimensionless}) and 
(\ref{masscons_dimensionless}) should coincide. In our standard 
model, we find that the mass distribution $m(x)$ from 
equation~(\ref{mass_dimensionless}) remains more or less steady in
time everywhere, whereas that from equation~(\ref{masscons_dimensionless})
shows large variations from one time to another inside a dimensionless 
radius of $\sim 0.3$. The variations are caused by unsteady mass 
accretion due to frequent reconnection events. The variations are 
much reduced if we average the mass distributions over timescales 
that are long compared with the durations of individual reconnection 
events but short compared with the total simulation time, which is
$\sim 1.6\times 10^{12}$~sec. 
The result is illustrated in Fig.~\ref{M_av}, for a representative 
period of time between $t=3.0\times 10^{11}$ and $7.5\times 10^{11}$~sec; 
the averaging is done over 90 outputs taken at intervals of $5\times 
10^9$~sec. Except for some wiggles within a small radius of $\sim
0.1$, the two 
independently determined average mass distributions agree rather 
well. The maximum discrepancy (of order $\sim 4\%$) occurs around 
$x\sim 0.5$, close to the outer edge of the pseudodisk. Perfect 
agreement is not to be expected, given the finite sizes of the inner 
and outer boundaries of the computation domain. The good agreement 
suggests that the collapse solution fluctuates around a well defined 
self-similar solution, with a mass distribution close to those 
shown in the figure. 

A similar closure relation can be derived for the angular momentum, 
allowing further verification of the approximate self similarity 
of this solution. We start from the equation for angular momentum 
conservation:
\begin{equation}
{\partial L(r,t)\over \partial t}=-\int \rho \varpi V_\varphi V_r dS
+{1\over 4\pi} \int B_\varphi B_r \varpi dS,
\label{angmom_cons}
\end{equation}
where $L(r,t)$ is the angular momentum enclosed within a sphere
of radius $r$ at a time $t$, and the two terms on the right
hand side are, respectively, the angular momentum advected into
the sphere by fluid motion per unit time and the magnetic torque
acting on the surface of the sphere, $S$. The quantity $\varpi$ 
is the cylindrical radius. If the collapse is 
strictly self-similar, the angular momentum can be written as
\begin{equation}
L(r,t)={a^5 t^2\over 2 G} l(x),
\label{angmom_scaling}
\end{equation}
where the dimensionless angular momentum $l(x)$ is given by 
\begin{equation}
l(x)=\int_0^x\int_0^\pi \alpha v_\varphi {\tilde x}^3 
\sin^2(\theta) d\theta d{\tilde x}. 
\label{angmom_dimensionless}
\end{equation}
Using equation~(\ref{angmom_scaling}), we can express the left-hand 
side of equation~(\ref{angmom_cons}) as 
\begin{equation}
{\partial L(r,t)\over \partial t} = {a^5 t\over 2G} \left[ 2 l(x)-x 
l^\prime(x) \right],
\label{chainrule_angmom}
\end{equation}
where $l^\prime=dl/dx$. Equating the right-hand side of 
equation~(\ref{chainrule_angmom}) with the right-hand side 
of equation~(\ref{angmom_cons}) yields
\begin{equation}
l(x) = {1\over 2} \left[ \int_0^\pi \alpha(x,\theta) 
v_\varphi 
x^4 \sin^2(\theta) d\theta +\int_0^\pi \alpha(x,\theta) v_\varphi 
x^3 \sin^2(\theta) (-v_r) d\theta + \int_0^\pi b_r b_\varphi x^3 
\sin^2(\theta) d\theta \right] 
\label{angmom_cons_dimensionless}
\end{equation}
which is the equation of angular momentum conservation in the 
self-similar space. Compared with the dimensionless equation 
for mass conservation (eq.~[\ref{masscons_dimensionless}]), 
there is an extra factor of $1/2$ on the right hand side 
(resulting from the quadratic rather than linear dependence 
of angular momentum on time) and an additional (third) term 
inside the square bracket that comes from magnetic braking. 
The first two terms have the usual interpretation: they 
are, respectively, the increase of angular momentum inside 
a sphere of constant dimensionless radius $x$ due to the 
expansion of the sphere and matter moving across the surface. 

We plot in Fig.~\ref{L_av} the averaged distributions of angular 
momentum computed from equations~(\ref{angmom_dimensionless}) and 
(\ref{angmom_cons_dimensionless}), over the same 90 outputs used
for Fig.~\ref{M_av}. The two distributions agree remarkably well, 
again indicating that the solution is nearly self-similar.

\subsection{Mass and Angular Momentum Distribution in Self Similar Space}

It is instructive to examine the dimensionless mass distribution 
$m(x)$ more closely. In Fig.~\ref{M_av}, we plotted the two terms 
on the right hand side of equation~(\ref{masscons_dimensionless}) 
separately. The contribution to $m(x)$ due to the advection term 
(proportional to 
the infall speed ${-v_r}$) has a minimum value of $\sim 0.5$ 
near the radius $x\sim 0.5$, where the infall is slowed down at the 
magnetic barrier. It increases inwards as the temporarily slowed 
down material picks up infall speed (see Fig.~\ref{vrvphi_eq}). The 
dip is compensated, to a large extent, by a peak in the contribution 
due to volume expansion (for a sphere of fixed dimensionless radius 
$x$). The peak is a result of the density enhancement in the
low-$v_r$, stagnation region.

Note that the dimensionless mass $m(x)$ approaches a finite value as 
$x\to 0$. The mass near the origin is simply the central point 
mass. It has a dimensionless value of 
$m_0 = 2G M_0/(a^3 t) \sim 3.6$. This mass is nearly twice the 
classical value for 
the inside-out collapse of a singular isothermal sphere (SIS), which 
has $m_0=1.95$ in our mass units \citep{1977ApJ...214..488S}. 
Part of the increase comes from the fact that the initial configuration 
for collapse is, on average, denser than the SIS. Another reason is 
that, unlike the SIS, the initial configuration is out of exact force 
balance, which induces a subsonic, global 
contraction. The mass accretion onto the point mass is hindered, on 
the other hand, by rotation, which tends to slow down the collapsing 
material through centrifugal force. Indeed, in the absence of a 
magnetic field, most of the collapsed material in the standard 
model would be rotationally supported rather than falling to the 
center. The relatively high rate of {\it central} mass accretion is made 
possible by continuous magnetic braking. 

To understand the effects of magnetic braking on the angular momentum 
evolution in more detail, we plot in Fig.~\ref{L_av} the individual 
terms on the right hand side of 
eq.~(\ref{angmom_cons_dimensionless}). 
As with the mass distribution $m(x)$, the 
contribution to the angular momentum $l(x)$ inside a sphere of 
radius $x$ due to advection (second term) shows a dip near 
$x\sim 0.5$, as a result of infall deceleration, whereas 
that due to volume expansion (first term) shows a peak in the 
same region, as a result of angular momentum pileup. A crucial 
difference is that, unlike $m(x)$, $l(x)$ does not asymptote to 
a finite value near the origin. Rather, it approaches zero at 
small radii. The implication is that, {\it although mass falls 
onto the central object at a relatively high rate, little 
angular momentum is accreted}. The angular momentum removal 
can be seen most clearly by comparing the angular momentum 
advected into a sphere by mass motion (second term) and that 
removed by magnetic braking from the same surface (third term). 
>From Fig.~\ref{L_av}, we find that these two terms nearly cancel 
each other over an extended region (up to $x\sim 0.4$). The 
near cancellation points to a detailed balance of inward and 
outward transport of angular momentum, which enables a large 
amount of matter to be accreted, but little angular momentum. 
The balance is at the heart of the magnetic braking-driven mass 
accretion. 

\subsection{Magnetic Bubble} 

The magnetic bubble plays a key role in the angular momentum 
redistribution. Angular momentum is transported outward in 
this region through both outflowing gas and magnetic torque. In 
the standard model, roughly half of the angular momentum advected 
across a given radius by the equatorial, infalling, pseudodisk 
is carried away by the outflowing material; the other half 
by the highly twisted magnetic field. It is the toroidal 
field that drives the expansion of the bubble, as in the so-called 
``magnetic tower'' investigated by \citet{2003MNRAS.341.1360L}.
The strength of the toroidal field is determined by the ambient pressure 
that confines the bubble in the lateral direction. In Fig.~\ref{pressures}, 
we plot the total pressure and
its various components as a function of angle at a representative
radius $x=0.3$ (roughly the middle radius of the bubble). The 
total pressure does not vary much at different angles, but it 
is dominated by different components in different regions. In the 
pseudodisk near the equator ($\theta=\pi/2$), it is dominated by 
the thermal pressure, because of high density. The pressure due 
to the toroidal magnetic field dominates at intermediate 
latitudes inside the bubble. It is more than an order of magnitude 
stronger than the pressure due to the poloidal field, and a factor 
of $\sim 2.5$ times larger than the thermal pressure. The bubble 
is therefore magnetically dominated, with a plasma-$\beta$ less 
than unity. The polar region is even more strongly dominated by 
the magnetic field, although by the poloidal rather than the 
toroidal component. 

The kinematics of the three regions are also quite distinct. Over 
most of the equatorial pseudodisk, the infall and rotation speeds 
are comparable. Both are supersonic and super Alfv\'enic. The 
infall dominates the rotation in the polar funnel. It is supersonic 
but sub-Alfv\'enic (because of a high Alfv\'en speed associated 
with the strong poloidal field and low density). The bubble, on the 
other hand, is a toroidally dominated structure in both magnetic 
field (by definition) and velocity field. The mass-weighted average
rotation speed inside the bulk of the bubble (within a dimensionless
radius $x=1$) is about $2.2~a$, whereas that for the average poloidal 
speed is slightly less than the sound speed $a$, which is smaller 
than the Alfv\'en speed. The meridian flow inside the bubble is 
thus generally sub-Alfv\'enic. It is in contact, through magnetosonic
waves, with the base of the bubble, where a toroidal magnetic 
flux is constantly generated out of the poloidal flux (dragged in
from large distances) by 
differential rotation. It is the continuous insertion of new 
toroidal flux into the bubble that drives it to expand \citep{2003MNRAS.341.1360L}. 

The sub-Alfv\`enic motions indicate that the bubble can be viewed
as being more or less quasi-static in the meridian plane. It is 
then natural to ask: what supports the bubble against the gravity 
(which comes mostly from the point mass at the radii of the bubble, 
with some contribution from the self-gravity)? It turns out that 
the support is provided mainly by the rotation and toroidal magnetic 
field; the latter acts on the bubble material through a combination 
of magnetic pressure gradient and tension force. In some sense, 
the slowly expanding bubble is a partly rotationally and partly 
magnetically supported structure that forms in place of the  
purely rotationally supported structure that would form in the 
absence of magnetic braking. It is a generic, potentially 
observable structure that we will discuss in some depth in 
\S~\ref{magnetogyrosphere}. 

The bubble is a depository of mass and angular momentum. 
The dimensionless mass and angular momentum are more or less 
constant in time, with small 
fluctuations due to reconnection events. The bubble has an 
average mass $m_{\rm b} \sim 2.3$ within a radius of $x=1$, which 
is about 64\% of the central point mass ($m_0\sim 3.6$). It 
carries a total dimensionless angular momentum of $\sim 1.4$. 
The average specific angular momentum is therefore $l_{\rm b}\sim 
0.61$. It corresponds to a centrifugal radius of 
\begin{equation} 
x_c\sim {2 l_{\rm b}^2 \over m_0} \sim 0.2. 
\label{centrifugalrad}
\end{equation} 
If a fraction of the rotating bubble material were to recollapse 
at a time $t$ in the absence of further magnetic braking, it would 
form a disk of size comparable to the centrifugal radius  
\begin{equation}
r_c \sim 10^3 \left({a\over 0.3 {\rm km/s}}\right) 
\left({t \over 10^5 {\rm yr}}\right)~{\rm AU}
\label{cenrad}
\end{equation}
around the central object. We will discuss the possibility of
late-time disk formation in \S~\ref{magnetogyrosphere}. 
  
Another way to gauge the dynamic state of the bubble is to consider 
its energies. The total (thermal, kinetic and magnetic) energy is 
close to the gravitational binding energy in the potential of the 
central point mass. It is well below the gravitational binding 
energy when the self-gravity of the core is also taken into account. 
The bulk of the bubble material cannot escape to infinity by itself. 
It is expanding slowly outward only because it is being pushed 
continuously from below, as more and more mass and toroidal magnetic 
flux being inserted into the bubble. If for some reason the 
insertion were to stop suddenly, the bubble (at least a 
portion of it) may recollapse to form a disk. The existence 
of a large amount of bound (bubble) material of high specific 
angular momentum not far from the center indicates that, {\it although
magnetic braking can allow mass to fall into the center at
a high rate, the overall angular momentum problem is not 
decisively resolved.}
 We will speculate in the discussion 
section implications of this situation, including the possible 
role of protostellar winds in dispersing away the bubble and 
the angular momentum accumulated in it.

\section{Effects of Magnetic Field Strength}
\label{fieldstrength}

Having examined in some detail a standard model with a specific 
mass-to-flux ratio of $\lambda=13.3$, we now turn our attention 
to cases with different degrees of magnetization. We will first 
concentrate on the relative weak field cases with $\lambda$ 
between 400 and 13.3 (\S~\ref{weakfield}). The strong field cases
with $\lambda=4$ and $8$ behave quite differently; they will 
be discussed separately in \S~\ref{strongfield}.  

\subsection{Relatively Weak Field Cases} 
\label{weakfield}

Before going into details of the various collapse solutions, we first
discuss their general behaviors based on the dimensionless central 
mass $m_0$. It is defined as $m_0=2 G M_0/(a^3 t)$, where $M_0$ is
the central mass. The dimensionless mass $m_0(t)$ at any given time is 
simply the averaged mass accretion rate up to that time, normalized by 
$a^3/(2G)$. It is plotted as a function of time in Fig.~\ref{Mstar_LowB} 
for all cases with $\lambda$ between 13.3 and 400. Since mass accretion 
is driven by magnetic braking, which in turn depends on the field 
strength, one expects $m_0$ to decrease as $\lambda$ increases. This 
trend is indeed seen in Fig.~\ref{Mstar_LowB}, except for the three
most weakly magnetized cases ($\lambda=400$, 200 and 133). For these 
extremely weak field cases, the trend is not as clear, especially at 
late times. It turns out that these are the cases where a 
rotationally-supported, equilibrium disk is present.  
They are discussed in \S~\ref{weakextreme}. Even the weakest field 
case has a significant mass accretion, however, at a rate well above 
that in the non-magnetic case ($\lambda=\infty$; the lowest curve in 
the figure) at late times. The initial burst of mass accretion in 
the non-magnetic case comes from the finite size of the inner hole 
and the presence of a small ``seed'' mass that was placed at the
center to induce collapse, rather than any angular momentum 
redistribution. 

The initial burst in mass accretion is also present in the magnetized 
cases, although it is relatively inconspicuous in the two strongest
field cases ($\lambda=13.3$ and $20$). In both cases, the (average) 
mass accretion rate quickly approaches a plateau, indicating that 
a nearly self-similar state is reached; the small oscillations are 
related to magnetic flux accumulation in the central hole (see  
\S~\ref{strongfield} below). Indeed, the collapse 
solution in the $\lambda=20$ case is very similar to that of the 
standard $\lambda=13.3$ case discussed in the last section. It will 
not be discussed separately further. 

The intermediate cases with $\lambda=40$ and 80 are particularly 
interesting. They represent the transition from the weaker field 
cases where a rotationally-supported, equilibrium disk is 
present 
to those where the disk formation is suppressed. These cases are 
discussed in \S~\ref{transition}. 

\subsubsection{Disk Formation in Extremely Weak Field Cases}
\label{weakextreme}

The rotating collapse is modified significantly by the magnetic field 
in the case of $\lambda=400$, which corresponds to a field strength 
of merely $0.25~\mu G$ at the typical core radius of 
0.05~pc. In the absence of any magnetic field, the initial 
configuration collapses into a self-gravitating 
ring (centered at a dimensionless radius $x=0.05$). The ring 
formation is suppressed by the weak field, and a disk is 
produced in its place, as shown in Fig.~\ref{snap_Bpt01}. The figure 
is a snapshot at a representative time $t=2.0\times 10^{12}$~sec, 
plotted in self-similar quantities. The snapshots at other times 
are similar to the one shown. Prominent in the figure is a flatten 
high-density disk, surrounded by a lower density, more puffed up 
``corona.'' Despite the weakness of the initial field, the 
magnetic pressure due to toroidal field actually dominates the 
thermal pressure, by a factor of $\sim 2.5$ on average. The 
inner part of the disk is more strongly magnetized, with the 
ratio of magnetic to thermal pressures greater than 10 inside 
a radius of $\sim 0.01$, perhaps as a result of a shorter orbital 
period, which allows the field lines to wind up more turns in a 
given time. The poloidal magnetic field inside the disk is, on 
the other hand, very weak, with an energy of order $0.1\%$ of the 
thermal energy on average. The weak poloidal field is susceptible 
to magnetorotational instability \citep[MRI;][]{1998RvMP...70....1B}. 
The role of MRI, as opposed to simple twisting of a pre-existing 
poloidal field, in redistributing angular momentum in the disk 
remains to be quantified (see discussion in \S~\ref{lowbetadisk} 
below). 

The disk is rotationally supported. The best evidence for rotational 
support comes from Fig.~\ref{Disk_speeds}, where the rotation speed 
of the material on the equator is compared to the equilibrium 
rotation speed computed from the (total) gravitational potential 
at the representative time shown in Fig.~\ref{snap_Bpt01}. 
Inside the disk (within a radius of about $0.1$), the 
actual rotation speed is close to, but slightly less than, the
equilibrium value. The small deviation is caused by additional
supports from thermal pressure and magnetic forces. Outside
the disk, the rotation speed is substantially less than the
equilibrium value. Also plotted in Fig.~\ref{Disk_speeds} is 
the radial component of the velocity of the material on the 
equator. It shows an outer region that infalls supersonically, 
and an inner region that is nearly static.  
The former is expected of the collapsing envelope, whereas the 
latter is consistent with a rotationally-supported disk 
that is close to an equilibrium in the radial direction. 
The disk does have a small, non-zero, radial velocity, which is 
responsible for transporting material from the outer disk to 
the inner disk. The 
transport, driven by angular momentum redistribution, enables 
mass accretion onto the central object, at 
a dimensionless rate $\sim 0.7$, which is about 1/3 of the 
rate for SIS \citep{1977ApJ...214..488S}.  

The rate of mass accretion from the disk to the central object is 
small compared to that from the envelope to the disk. As a result,
the bulk of the material collapsed from the envelope 
is stored in the disk rather 
than the central object. The relative smallness of the central mass 
is also evident from the equilibrium rotation curve shown in 
Fig.~\ref{Disk_speeds}, which deviates significantly from a Keplerian
curve, especially in the outer part of the disk. The deviation, 
particularly the bump near the disk edge around $x\sim 0.1$, is 
an indication that self-gravity is important; indeed, it dominates 
the gravity of the central mass beyond a dimensionless radius 
$\sim 0.04$.  

The $\lambda=200$ and 133 cases are qualitatively similar to the
$\lambda=400$ case. They all produce a persistent, 
rotationally supported disk. There are some quantitative 
differences, however. The  
magnetic barrier, barely noticeable in the weakest field 
case, are rather prominent in the stronger field cases. Also, 
as the field strength increases, the disk 
becomes more dynamically active in the radial direction. The 
infall speed can occasionally become supersonic in some (usually 
narrow) regions, indicating that sporadic disk disruption has 
begun already in localized places, especially in the $\lambda=133$ 
case. The disruption becomes more widespread and persistent as the 
field strength increases further. 

\subsubsection{Transition from Disk Formation to Suppression}
\label{transition}

Sustained disk disruption happens in the case of $\lambda=80$. The 
disruption is illustrated in Fig.~\ref{vr_specmom_Bpt05}, where 
the radial component of the velocity, $v_r$, on the equator 
is plotted at a representative time $t=2\times 10^{12}$~sec. Unlike 
the weaker magnetic field case of $\lambda=400$ shown in 
Fig.~\ref{Disk_speeds}, where the radial velocity nearly vanishes 
everywhere in the disk, $v_r$ is close to zero in four discrete
(narrow) regions, with rapid, transonic or even supersonic, infall 
in between. 

The stagnation point at the largest radius is the magnetic barrier,
as in the standard, stronger field case (see Fig.~\ref{vrvphi_eq}).   
In the standard case, there is a second stagnation point interior to 
the magnetic barrier, which we have interpreted as the centrifugal 
barrier, where the infall is slowed down temporarily and the magnetic
braking is locally enhanced. In the 
$\lambda=80$ case, the stagnation points interior to the magnetic
barrier are also centrifugal barriers. The reason for the 
existence of more than one barriers can be seen from the upper curve in 
Fig.~\ref{vr_specmom_Bpt05}, where the specific angular momentum
is plotted as a function of radius. Even though a good fraction (about
half) of the 
angular momentum is removed at the first centrifugal barrier, there
is still a significant amount left interior to it. The material  
recollapses quickly towards the center, spinning up at the same 
time, until a second centrifugal barrier is encountered. The slowdown  
allows more time for the fluid rotation to twist up the field lines. 
The magnetic braking removes enough angular momentum from the material in 
the equatorial region to enable it to recollapse for a third time, 
all the way to near the inner boundary, where the infall is 
rearrested by rotation. In some sense, the rotating material tries 
to settle into a rotationally supported disk, but the attempt is 
repeatedly frustrated by magnetic braking. The result is a collection 
of discrete rings that are temporarily supported by the 
centrifugal force, separated by regions of rapid collapse, where 
the rotational support is insufficient to balance the gravity. 
Note that in 3D, the rings may break up into arcs or clumps, some of which may 
become self-gravitating and form (sub-)stellar objects.

The material in the stagnation region near a centrifugal barrier
loses a large fraction of its angular momentum, and collapses 
quickly inwards after exiting the region. Why is it possible to 
remove a large fraction of the angular momentum before the material 
begins collapsing dynamically? One may expect the gas to move 
(inward) out of the stagnation region as soon as its specific
angular momentum 
is reduced slightly below the local equilibrium value, because of an 
imbalance of gravity and centrifugal force. If this were true, the
amount of angular momentum reduction achievable in the region 
will be quite limited. 
However, it takes a finite amount of time for even the full gravity 
to reaccelerate the material to (say) the sound speed from at rest, no 
matter how small a value its angular momentum has been reduced to; 
in many cases, the gravity must first reverse the (slowly) expanding 
motion often found near the centrifugal barrier, which takes
additional time. If 
the magnetic torque is strong enough, it can remove a large fraction, 
if not all, of the angular momentum of the gas within that time. 
In other words, it is the inertia of the stalled gas that enables the 
magnetic braking to reduce the specific angular momentum significantly 
below the local equilibrium value. The efficient, localized, braking 
is responsible for the unique, ``hopping,'' pattern of alternating 
stagnation and rapid infall seen in Fig.~\ref{vr_specmom_Bpt05}.

The rotationally supported rings show up clearly in the density map 
(Fig.~\ref{disk_unitvec_Bpt05}). They are overdense regions near 
the equator created by stalled infall, 
which leads to a pileup of matter.
The slowdown also allows more time for magnetic braking to operate. 
The clearest evidence for enhanced braking comes from the outer most 
ring near the dimensionless radius $x\sim 0.1$: there is an outflow 
driven from that location at an acute angle with respect to the 
equatorial plane, as in the magnetic bubble of the standard case. 
Evidence for braking-driven outflows from the inner two rings is 
less clean-cut; they may be masked by the chaotic meridianal flow 
pattern in the puffed-up region outside the equatorial rings. The 
chaotic motions are generally subsonic. They are dominated in 3D by 
supersonic rotation. Indeed, the puffed-up structure is equivalent 
to the magnetic bubble in the standard model: both are supported
by a combination of (mostly toroidal) magnetic fields and rotation.
In some sense, the puffed-up structure is a collection of (relatively 
weak) bubbles, driven from multiple centrifugal barriers whose 
locations vary in time. Such a structure is also present in the 
somewhat stronger field case of $\lambda=40$. We believe it is a 
generic feature of the magnetized rotating collapse when the field
is strong enough to disrupt the rotationally supported disk but 
too weak to channel the outflowing material into a single, coherent
bubble. This feature is discussed further in
\S~\ref{magnetogyrosphere}. 

We note that the number of discrete centrifugal barriers depends on 
the field strength. As the field strength decreases, more and more 
centrifugal barriers appear, until they start to overlap and merge 
into a single, more or less contiguous structure --- a rotationally 
supported disk (see the $\lambda=400$ case discussed above). 
Conversely, the disk is replaced by fewer and fewer centrifugal 
barriers as the field strength increases. For the standard, 
$\lambda=13.3$ case, the number of identifiable centrifugal barriers 
is one at most times. This number can drop to zero for more 
strongly magnetized cases, to which we turn our attention next. 

\subsection{Relatively Strong Field Cases} 
\label{strongfield}

We first focus on the time evolution of the central point mass 
(Fig.~\ref{Mstar_HighB_noscaled}), which provides an overall  
measure of the collapse solutions. The point mass at a give time 
remains similar or even becomes smaller as the field strength 
increases, unlike the moderately weak field ($\lambda=80$ to 13.3) 
cases. The trend is especially clear for the strongest field  
case of the whole set ($\lambda=4$). This change in the dependence 
of point mass on the field strength indicates a change in the 
solution behavior. In the weak field cases, the mass accretion 
is limited by angular 
momentum removal, whose rate generally increases with the field strength. 
Angular momentum removal is no longer the limiting factor in the 
strong field cases: the (poloidal) fields at small radii are 
apparently strong enough, especially after amplification from 
collapse, that a relatively small twist can remove most of the 
angular momentum of the collapsing material and redistribute it 
to large distances. The limiting factor shifts instead to the 
magnetic tension, which replaces the rotation as the main 
impediment to mass accretion. The tension is stronger (relative 
to gravity) for a smaller mass-to-flux ratio $\lambda$. Indeed, 
if $\lambda < 1$ (i.e., a magnetically subcritical core), the 
tension force would prevent core collapse and mass accretion 
altogether. 

Unlike angular momentum, the magnetic flux cannot be removed in 
the ideal MHD limit that we are working in. It accumulates near 
the central object --- in the central hole in our simulation. 
The flux accumulation and its sudden release through reconnection 
allow the magnetic tension to suppress mass accretion onto the
central object over an extended period of time even in a 
magnetically supercritical core with an $\lambda$ significantly 
larger than unity. They are responsible for 
the step-like variations of the point mass shown in 
Fig.~\ref{Mstar_HighB_noscaled}. Before discussing magnetic 
reconnection in \S~\ref{reconnection}, we first consider the 
collapse solution at a time representative of the conditions 
before a reconnection event.  

\subsubsection{Removal of Angular Momentum in a Magnetic Braking-Driven Wind} 
\label{nodisk}

We show in Fig.~\ref{snap_B1pt0} a snapshot of the $\lambda=4$ case 
at a representative time before reconnection. Although one can still
divide the collapse solution into four regions as in the standard 
case (collapsing envelope, equatorial pseudodisk, magnetic bubble 
and polar funnel, see Fig.~\ref{snap}), these regions are less
distinct. In particular, the magnetic bubble (defined as the 
region where the toroidal field is stronger than the poloidal field) 
is much less prominent than in the standard $\lambda=13.3$ case. 
It is located at a larger distance from the center and contains a 
much lower mass and angular momentum. Within a radius of $x=3$, the 
dimensionless mass of the bubble is only 0.18 and angular momentum 
is 0.28. These are to be compared with the bubble mass of $\sim 2.3$ 
and angular momentum of $\sim 1.4$ estimated earlier for the standard 
case. Indeed, the bubble can be more profitably viewed as part of 
a broader magnetic braking-driven wind, where the field lines are
twisted to varying degrees. The (poloidal) field is strong enough 
that even a relatively small twist can exert an appreciable torque 
on the gas: for a fixed twist angle, the magnetic torque is 
proportional to the field strength squared. Compared to the
standard case, the twisted field lines unwind faster because of 
a higher Alfv\'en speed. Another difference is the absence of a 
prominent magnetic wall in which the field lines are severely 
compressed; the stronger fields are more difficult to compress. 
As a result, the collapsing envelope merges more smoothly into 
the equatorial pseudodisk, which is now much thicker. The 
centrifugal force inside the pseudodisk remains substantially 
below the gravity in general. In some sense, parcels of infalling 
material are chasing their centrifugal barriers but unable to 
reach them because of efficient magnetic braking, which continuously 
moves the barriers inward. We conclude that the formation of a 
rotationally supported disk is completely suppressed in the 
relatively strong field regime, in agreement with the earlier 
work of ALS03. 

\subsubsection{Magnetic Reconnection and Episodic Mass Accretion}
\label{reconnection}

A feature not apparent in the early work is episodic mass accretion. As 
can be seen from Fig.~\ref{Mstar_HighB_noscaled}, most of the central 
mass in the $\lambda=4$ case is accreted in discrete bursts, which 
are caused by magnetic reconnection. To determine how the bursts are
triggered, we focus on the large burst around $t=9.1\times 10^{11}$~sec. 
A snapshot of the inner part of the collapse solution before 
reconnection is shown in the left panel of Fig.~\ref{burst}. Note 
the nearly radial field lines that thread the surface of the inner 
hole. Together with their mirror images (of opposite polarity) 
in the lower quadrant (not shown), they form a split magnetic
monopole. The 
monopole compresses the material in the equatorial region into a 
thin, dense (pseudo)disk, which collapse supersonically. As more 
and more disk material falls into the point mass, more and more 
field lines are added to the central hole. The growing monopole 
pushes on the disk material harder and harder from above and 
below. When field lines of opposite polarities are pressed into 
a single cell on the equator, they reconnect quickly due to 
numerical diffusion. We observe a maximum in the magnetic 
flux that threads the inner hole shortly before the onset of violent 
reconnection. There is a brief period of time when matter falls 
into the central hole but not the field lines. The separation 
indicates that significant magnetic diffusion is present before 
the violent reconnection. The reconnected field lines pull the 
disk material outward (see the right panel of Fig.~\ref{burst}), 
leaving behind an evacuated region that is unable to keep the 
oppositely directing, monopolar field lines apart. As a result, 
the field lines threading the central hole rush onto the disk, 
adding to the reconnected field lines that accelerate the 
equatorial material outward. Except for a tiny amount of mass 
falling into the central hole in the polar region, mass accretion 
is essentially terminated. 

As the dense blob (or ring in 3D) shown in the right panel of 
Fig.~\ref{burst} moves outwards, it gradually slows down for several 
reasons. First, it ploughs into infalling material, which cancels 
some of its outward momentum. Second, as the mass of the blob 
increases, through both sweeping up material along its path and 
accretion from above and below along field lines, the gravitational 
pull of the central point mass increases as well. In addition, 
the magnetic tension that pushes the blob outwards from behind 
weakens with time, as the bent field lines straighten up. Sooner 
or later, there will be enough material accumulated for the 
gravity to overwhelm the magnetic forces and pull the matter 
toward the center again. The recollapsing flow in the equatorial 
region typically moves at a speed well below the local free fall 
speed, because it has to drag along with it a rather strong 
magnetic field, which resists the gravitational pull through 
tension force. 

Once the bulk of the recollapsing material reaches the inner hole,
it produces a burst in mass accretion. At the same time, the 
accreted material deposits its magnetic flux on the surface of the
inner hole, creating a growing split magnetic monopole that, in 
time, would shut off the mass accretion through reconnection, and 
eject the remaining material to larger distances, starting another
cycle of mass accumulation, collapse, and ejection. In some sense, 
the split monopole acts as an ever tightening ``magnetic gate:'' 
as magnetic flux accumulates, mass 
accretion becomes more and more difficult, until shut off almost 
completely by reconnection; the reconnection allows a large amount of 
magnetic flux (associated with the central point mass) to act on a 
small amount of (disk) matter. 

In summary, we have computed the collapse of rotating, strongly 
magnetized cores with much higher angular resolution at small radii 
than the previous calculations that employ a cylindrical coordinate 
system. We confirmed the conclusion of ALS03 that the magnetic 
braking in such cores is too strong for the rotationally supported 
disks to form in the ideal MHD limit. We went one step further, 
showing that the disk suppression holds true even in the presence 
of (numerically triggered) reconnection. The main effect of the 
reconnection events is to generate episodic mass accretion. 

\section{Effects of Rotation Speed} 
\label{rotation}

The initial rotation speed $v_0=0.5$ (in units of sound speed) 
adopted in the standard 
simulation is on the high side of the observationally inferred 
range. Here we discuss three variants 
of the standard model, with $v_0=0$, $0.125$ and $0.25$, 
respectively. Without rotation, the collapsing envelope is
expected to fall into the central object at a high rate. This 
is indeed the case, as shown in Fig.~\ref{Mstar_diffrot}. The 
dimensionless point mass quickly approaches a nearly constant
value, indicating that the non-rotating collapse is approximately
self-similar. There are small oscillations in the point mass, 
which result from magnetic flux accumulation at the inner hole, 
as in the standard simulation. This episodic behavior is therefore
independent of rotation, in accordance with the interpretation
based on reconnection that we advanced in the last section. The 
value of the dimensionless point mass, $\sim 8$, corresponds to 
a mass accretion rate that
is about 4 times higher than the rate for the collapse of a 
static SIS \citep[1.95;][]{1977ApJ...214..488S}. 
The accretion rate decreases as the 
rotation speed increases, as shown in Fig.~\ref{Mstar_diffrot}. 
For the relatively slowly rotating cases of $v_0=0.125$ and $0.25$, 
the point mass started out as high as that in the non-rotating
case, but gradually approaches a lower, nearly constant value,
indicating that an approximate self-similarity is reached in
these cases as well. 

To illustrate the effects of rotation on the structure of the 
magnetized collapsing flow, we show in Fig.~\ref{multi_diffrot} 
a representative snapshot for each of the cases with $v_0=0$,
$0.125$, $0.25$ and $0.5$. In the absence of 
rotation, three of the four dynamically distinct regions that we
have identified in the standard (rotating) model are clearly 
present: the polar funnel flow (Region IV), equatorial pseudodisk 
(region II), and collapsing 
envelope (Region I). They are separated by a magnetic wall, 
which divides the outer collapsing envelope from the inner 
two regions, both of which are 
dominated by the central split magnetic monopole.  
This magnetic structure is remarkably similar to
the one sketched in Fig.~1 of \citet{2006ApJ...647..374G}. Rotation 
changes the appearance of the collapse 
solution by inflating a bubble in between the polar funnel and
the pseudodisk. It is bounded from outside by the envelope. 
 As one may expect intuitively, the size 
(mostly the width) of the bubble increases with the initial 
rotation speed; for the same initial magnetic field, a faster
rotation should wind up the field lines more, generating a 
larger magnetic bubble, as shown in the figure. The difference
in the rotation speed does not appear to change the dynamics of 
the core collapse and magnetic braking fundamentally, however. 

\section{Discussion}
\label{discussion}

\subsection{Characteristic Mass-to-Flux Ratio for Significant Magnetic
  Braking}
\label{lambdabraking}

Our most important result is that the rotationally supported disk
can be disrupted by a weak magnetic field corresponding to a
mass-to-flux ratio $\lambda \gg 1$. This result may seem surprising 
at the first 
sight, in view of the general expectation that the field is 
dynamically important only when $\lambda \lesssim 1$. To illuminate 
the reason behind this seemingly unexpected result, we provide an 
order-of-magnitude estimate of the characteristic mass-to-flux 
ratio for significant magnetic braking from the following equation:
\begin{equation}
M_d l_d \sim {\vert B_{\varphi}\vert B_z \varpi_d \over 4\pi}\ 
(2 \pi \varpi_d^2)\ t,
\label{LeqTaut}
\end{equation}
where $M_d$, $l_d$ and $\varpi_d$ are the mass, specific angular
momentum and radius of the disk in the absence of magnetic 
braking at time $t$, and $B_\varphi$ and 
$B_z$ are the toroidal and vertical components of the magnetic field
on the disk surface. The equation is simply a condition on the rate 
of magnetic braking that would be required to remove all of the disk 
angular momentum within the time $t$. 

The disk mass $M_d$ is related to the magnetic flux threading the
disk, $\pi \varpi_d^2 B_z$, through the dimensional mass-to-flux 
ratio $\lambda/(2\pi G^{1/2})$, which is conserved along any field 
line during the core collapse in the ideal MHD limit. Eliminating
$B_z$ in terms of $M_d$ and $\lambda$ from equation (\ref{LeqTaut}),
we have 
\begin{equation}  
\lambda^2 \sim 2 \left( \frac{\vert B_{\varphi}\vert}{B_z}\right) 
\left(\frac{G M_d t}{l_d \varpi_d}\right). 
\label{lambda1}
\end{equation}
To simplify the right hand side of the equation further, we assume 
that the disk is self-gravitating and rotationally supported 
(as in the $\lambda=400$ case shown in Fig.~[\ref{snap_Bpt01}]), with 
a rotation speed $v_d\sim (G M_d/\varpi_d)^{1/2}$ and specific 
angular momentum $l_d\sim v_d \varpi_d$. The above equation then 
becomes
\begin{equation}  
\lambda^2 \sim 4\pi \left( \frac{\vert B_{\varphi}\vert}{B_z}\right)
\left(\frac{t}{p}\right), 
\label{lambda2}
\end{equation}
where $p= 2\pi \varpi_d/v_d$ is the disk rotation period. 

Clearly, the characteristic mass-to-flux ratio can greatly exceed 
unity for two reasons: (1) tight wrapping of the disk magnetic 
field, $\vert B_\varphi\vert \gg B_z$, which increases the rate of 
angular momentum removal through magnetic braking ($\propto \vert 
B_\varphi\vert  
B_z$), and (2) long braking time compared to the disk rotation 
period, $t \gg p$. The braking time $t$ can be identified with 
the time for core collapse. 

The two factors are not entirely independent of each other. Since 
the toroidal magnetic field is created out of the poloidal field
by rotation, the field line is expected to become increasingly more 
twisted with time. Indeed, it may not be unreasonable to 
approximate the magnetic twist by the number of turns that the 
disk has rotated around the center within time $t$, so that   
\begin{equation}
{{\vert B_\varphi\vert} \over B_z}\sim {t\over p}.
\end{equation}
In this case, the characteristic mass-to-flux ratio reduces to
\begin{equation}  
\lambda \sim 2\sqrt{\pi} \left(\frac{t}{p}\right).  
\label{lambda3}
\end{equation}

To obtain concrete numbers, let us consider a disk formed at time
$t$ out of a singular isothermal toroid with an (angle-averaged) 
initial mass distribution $M(r)= 2(1+H_0) a^2 r/G$ and spatially constant 
rotation speed $V_0$. Suppose, as in the case of SIS \citep{1977ApJ...214..488S}, 
half of the mass within a radius $r= a t$ falls into the 
collapsed object, the disk, at any given time $t$. In this case, 
the disk mass $M_d\sim (1+H_0) a^3 t/G$ and specific angular
momentum $l_d \sim V_0 a t/2$, yielding a disk radius
\begin{equation}
\varpi_d \sim {l_d^2 \over G M_d}\sim {V_0^2 t\over 8 (1+H_0) a},
\end{equation}
and rotation period
\begin{equation}
p \sim {\pi \over 4(1+H_0)^2}\left({V_0\over a}\right)^3 t.
\label{period}
\end{equation}
Substituting equation~(\ref{period}) into equation~(\ref{lambda3}),
we finally have 
\begin{equation}
\lambda \sim {8(1+H_0)^2\over \sqrt{\pi}} \left({a\over V_0}\right)^3.   
\label{lambda4}
\end{equation} 
 
For the initial mass distribution with $H_0=0.4$ and rotation speed
$V_0=0.5 a$ adopted in \S~\ref{fieldstrength}, we obtain a 
characteristic mass-to-flux ratio for significant magnetic braking 
of $\lambda \sim 70$. This number is remarkably close to our 
numerically determined value for the disk disruption to begin, 
which is around 100. Given the crudeness of the estimate, the
agreement is encouraging. The analytic estimate brings out the 
key ingredient for efficient disk braking by even weak magnetic
fields: short disk rotation period compared 
to the core collapse time. In addition, it indicates that the disk 
in a more slowly rotating core can be disrupted by a weaker magnetic 
field, which makes physical sense. 

Besides the rotation rate, another factor that may also affect 
the value of the mass-to-flux $\lambda$ for significant magnetic 
braking is (non-axisymmetric) gravitational torque. The torque can 
speed up mass accretion through a self-gravitating disk, shortening 
the time available for field line wrapping and magnetic braking. 
This additional means of angular momentum redistribution is expected 
to decrease the efficiency of magnetic braking somewhat. To fully 
quantify this effect, high resolution 3D simulations are required. 
Our 2D (axisymmetric) calculations, while idealized in several ways 
(see \S~\ref{future} below), allow us to concentrate on the role  
of magnetic braking in disk dynamics without extra complications
from gravitational torques and fragmentation. We believe the basic 
conclusion that even weak magnetic fields with $\lambda \gg 1$ are 
dynamically important in disk formation and evolution is robust.

\subsection{Magnetogyrosphere}
\label{magnetogyrosphere}

Surrounding the strongly braked disk is a vertically extended 
structure where the bulk of the angular momentum of the material 
collapsed from the envelope is parked. The structure is supported 
by a combination of (toroidal) magnetic field and rotation. We term 
it ``magnetogyrosphere,'' to distinguish it from the purely 
magnetically dominated ``magnetosphere'' or the rotationally supported 
disk.

An example of the magnetogyrosphere is shown in 
Fig.~\ref{disk_unitvec_Bpt05}, 
where $\lambda=80$. In this particular case, the collapsing envelope material 
is channeled by the magnetic wall surrounding the magnetogyrosphere into 
the equatorial region, where the infalling material crosses multiple 
centrifugal barriers on its way to the center. The braking of the equatorial
material, 
particularly at the multiple centrifugal barriers, is what powers the rather 
chaotic motions inside the slowly expanding 
magnetogyrosphere. The dimensionless size of the magnetogyrosphere is $\sim 
0.25$, corresponding to a physical size of $\sim 10^3$~AU at a time of 
$2\times 10^{12}$~sec, for a sound speed of $a=0.3$~km/s. By this time, 
the central point mass is about $0.3$~M$_\odot$, whereas the mass of 
the magnetogyrosphere is about $1.1$~M$_\odot$ within $10^3$~AU of the 
origin (we ignore the rotating, outflowing region near the axis at 
larger distances that contains relatively little mass). Most of the 
material collapsed from the envelope is therefore stored in the 
magnetogyrosphere, 
rather than going into the central object. It is prevented from falling 
into the center by the combined effect of magnetic field and
rotation. To be quantitative, 
the mass weighted flow speed in the poloidal plane is only $0.13$~km/s 
inside the magnetogyrosphere, which is less than half of the sound 
speed $a=0.3$~km/s. This is consistent with the fact that the size 
of the magnetogyrosphere is increasing slowly, at a rate of roughly a 
quarter of the sound speed. For comparison, the free fall speed $V_{\rm 
ff}=[2 G M(r)/r]^{1/2}$ at the outer edge of the magnetogyrosphere 
at $10^3$~AU is $1.55$~km/s, an order of magnitude higher than the 
average poloidal speed. The bulk of the magnetogyrospheric material 
is levitated against the gravity by a combination of (mostly 
magnetic) pressure gradient and centrifugal force from rotation; 
the mass averaged rotation speed is $1.23$~km/s, not far from the 
free fall speed. 

Qualitatively similar behaviors are found in other moderately weakly
magnetized models of $\lambda=40$, 20, and 13.3. For 
example, in the standard model with $\lambda=13.3$, the magnetic 
braking is strong enough that only a single centrifugal barrier 
exists. After exiting the barrier, the equatorial material falls 
straight to the center whereas the extracted angular momentum is 
stored in a slowly expanding bubble --- the magnetogyrosphere, an 
extended structure again dominated by a combination of magnetic 
field and rotation (see \S~\ref{ss}). Compared with the $\lambda
=80$ case, the magnetogyrosphere contains less material (with a 
mass somewhat less than, rather than well exceeding, the central 
mass), which moves faster on average and in a more ordered fashion 
in the poloidal plane. 

To illustrate what the magnetogyrosphere may look like observationally, 
we show in Fig.~\ref{columndensity} the column density distributions 
of the $\lambda=80$ and $13.3$ viewed perpendicular to the axis. In both
cases, there is a highly-flattened, dense equatorial region, sandwiched 
by a more puffed-up structure. The former may potentially be mistaken 
for a rotationally-supported equilibrium disk around the central object;  
in reality, it is a highly dynamic structure that collapses 
supersonically and rotates significantly below the local equilibrium 
rate at most radii, except near the centrifugal barrier(s). The latter 
is the slowly expanding magnetogyrosphere. It is held up mostly by the 
(toroidal) magnetic field in the vertical direction and laterally by 
rotation.

\subsubsection{Connection to Observations: IRAM 04191 and HH 211}

Magnetogyrospheres should exist in some form. Fundamentally, this 
is because a rotating, collapsing core will necessarily develop a 
differential rotation, which inevitably interacts strongly with a 
poloidal magnetic field, as long as the matter and field are well 
coupled. The interaction is especially strong near the centrifugal 
barrier, where the slowdown of infall motion allows more time for 
the differential rotation to wrap up the field lines. When the field 
is strong enough to remove angular momentum from the collapsing 
material, but too weak to transport most of the removed angular 
momentum out of the system, a magnetogyrosphere develops. In 
principle, the magnetogyrosphere can be distinguished easily 
from the collapsing (inner) envelope, since both the velocity 
and magnetic field are dominated by the toroidal component in 
the former and by the poloidal component in the 
latter. They also have different volume and column densities, which 
may also help separating the two. In practice, the mass distribution 
and especially the kinematics of both the magnetogyrosphere and the 
inner envelope are expected to be modified by the powerful 
protostellar winds, which sweep the ambient medium into bipolar 
molecular outflows. To search for observational evidence for the 
magnetogyrosphere, the best place to start may be those Class 
0 sources that have relatively narrow molecular outflows. 

One of the best studied Class 0 sources is IRAM 04191. The source 
has a low luminosity of $\sim 0.1 ~L_\odot$ \citep*{1999ApJ...513L..57A,2006ApJ...651..945D}, 
which points to a central
stellar object of mass perhaps no more than 0.1~$M_\odot$. The 
object is surrounded by an extended envelope of mass $\sim ~0.5 
M_\odot$ within a radius of 4200~AU, estimated from dust continuum 
emission \citep{1999ApJ...513L..57A}. Molecular line observations have 
established that the envelope is both infalling and 
differentially rotating \citep*{2002A&A...393..927B}. It is well traced 
by $N_2H^+$, except close to the center, where the molecule 
appears to be depleted \citep{2004A&A...419L..35B}. \citet*{2005ApJ...619..948L} 
found that the $N_2H^+$ envelope consists of two kinematically 
distinct parts: a fluffy outer region and a dense inner structure
that resembles a thick, clumpy, tilted ring. The thick ring-like 
structure has an average radius $r_0 \sim 1400~AU$ (or $10^{\prime
\prime}$). The most puzzling, and potentially the most revealing, 
kinematic feature is the inferred infall speed at the radius $r_0$: 
$v_r (r\sim r_0) \sim 0$. The small radial speed implies a pileup 
of infall material, and may have created the dense ring in the 
first place. The stagnation of infall, if confirmed, would have 
strong implications for the dynamics of mass accretion and angular 
momentum evolution. It may mark the transition from the collapsing 
envelope to a circumstellar structure dominated by magnetic fields 
and rotation --- the magnetogyrosphere in our picture.

Rotation alone can in principle stop the observed infalling envelope at 
the radius $r_0$, as would be the case near a centrifugal barrier. 
The rotation speed at $r_0$ is estimated to be $V_{\varphi,0}\sim 
0.16$~km/s, corresponding to a centrifugal force $V_{\varphi,0}^2 / 
r_0 \sim 1.22\times 10^{-8}$~cm/s$^2$ per unit mass. It is to be 
compared with the gravitational acceleration $G [M_{\rm env}(r_0)+M_*]
/r_0^2\sim 7.56 \times 10^{-8}$~cm/s$^2$, where we have assumed 
0.1~M$_\odot$ for the central stellar mass and 0.15~M$_\odot$ for the 
mass within $r_0=1400$~AU (out of the estimated $\sim 0.5$~M$_\odot$ 
within 4200~AU). The ratio of gravitational to centrifugal forces is 
therefore $\sim 6$, although it may be lower by perhaps up to a factor 
of 2 due to uncertainties in the mass estimates. In any case, the 
inferred rotation does not appear to be faster enough to counter the 
gravity \citep{2005ApJ...619..948L}, which implies that the stagnation 
point $r_0$ is not a centrifugal barrier. This result is consistent with 
the constraint that an optically thick disk, if exists at all, must 
be smaller than 10~AU \citep{2002A&A...393..927B}, well inside the 
stagnation radius $r_0$. 

\begin{figure}
\epsscale{1.0}
\plotone{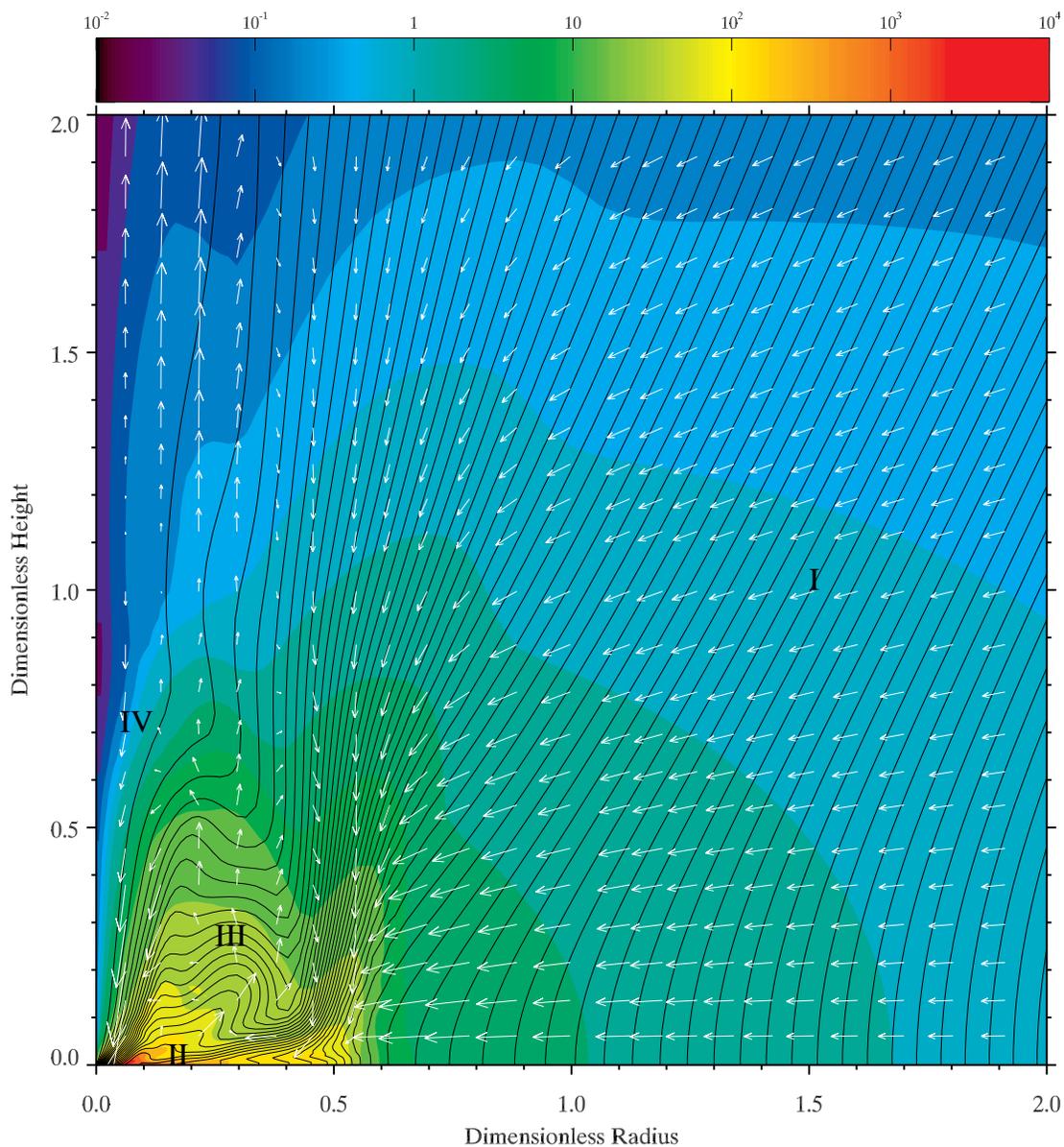}
\caption{Snapshot of the standard collapse solution at a representative
time. Shown in color is the distribution of dimensionless density in
the meridian plane, with the lines of 
magnetic field and velocity vectors superposed. 
The four dynamically distinct regions are labeled, corresponding to 
the collapsing envelope (Region I), equatorial pseudodisk (II), magnetic 
bubble (III) and polar funnel (IV). See the text for discussion.} 
\label{snap}
\end{figure}

\begin{figure}
\epsscale{1.0}
\plotone{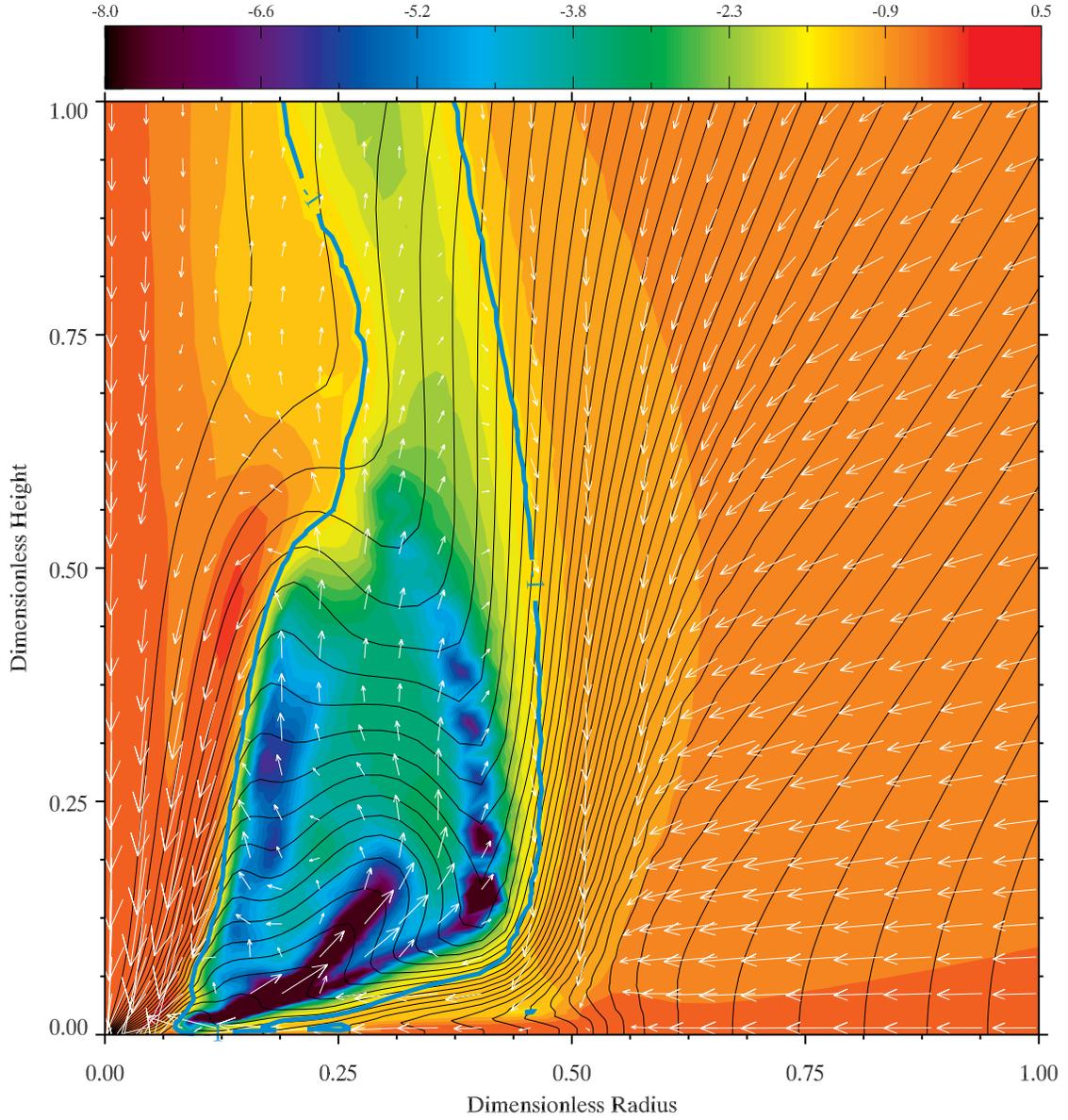}
\caption{Map of the magnetic twist. Superposed on the map are the 
lines of magnetic field (black) and 
velocity vectors (arrows). The toroidally dominated magnetic
bubble (enclosed within the blue contour labeled with ``-1'') 
is surrounded by the polar funnel flow to the left, collapsing
envelope to the right, and equatorial pseudodisk from below. } 
\label{twist}
\end{figure}

\begin{figure}
\epsscale{0.75}
\plotone{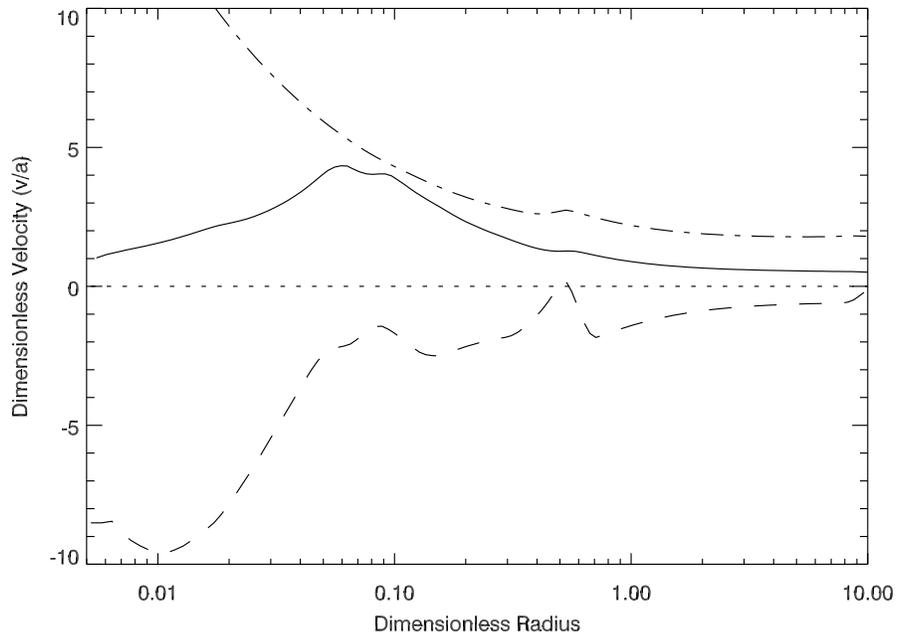}
\caption{Velocity distribution of the standard model on the equator. 
Plotted are the radial (dashed), rotational (solid) speeds of the 
fluid, and the rotation speed needed for support against gravity 
(dot-dashed).} 
\label{vrvphi_eq}
\end{figure}

\begin{figure}
\epsscale{0.75}
\plotone{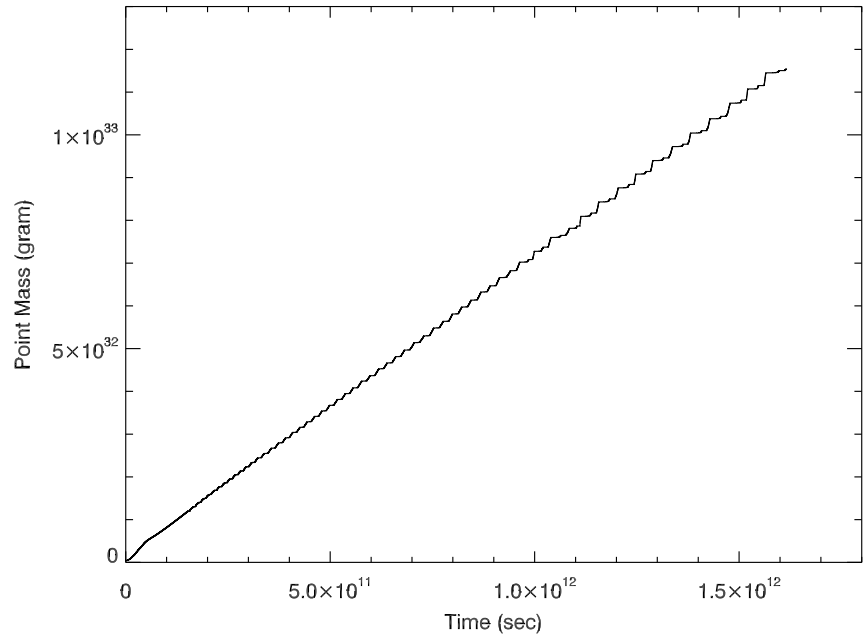}
\caption{Evolution of the central point mass for the standard model. 
The nearly linear increase of the mass with time indicates that the 
collapse solution is approximately self-similar. } 
\label{pointmass}
\end{figure}

\begin{figure}
\epsscale{0.75}
\plotone{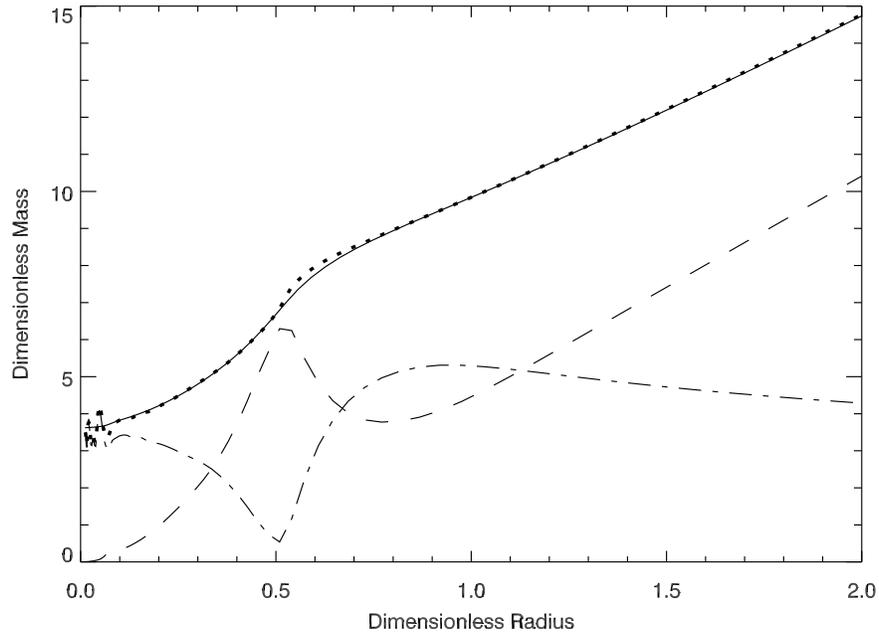}
\caption{Time averaged distributions of the dimensionless mass $m(x)$
  enclosed within an expanding sphere of fixed dimensionless radius 
$x$, computed in two different ways, using  
eqs~(\ref{mass_dimensionless}) (solid line) and
(\ref{masscons_dimensionless}) (dotted). The averaging is 
done between $3.0\times 10^{11}$ and $7.5\times 10^{11}$~sec. The 
agreement between the two curves is an indication that the collapse 
solution is approximately self-similar. Also 
plotted are the two terms on the right hand side of
eq.~(\ref{masscons_dimensionless}), showing the mass change   
due to volume expansion (dashed line) and mass crossing the 
sphere (dot-dashed).
}
\label{M_av}
\end{figure}

\begin{figure}
\epsscale{0.75}
\plotone{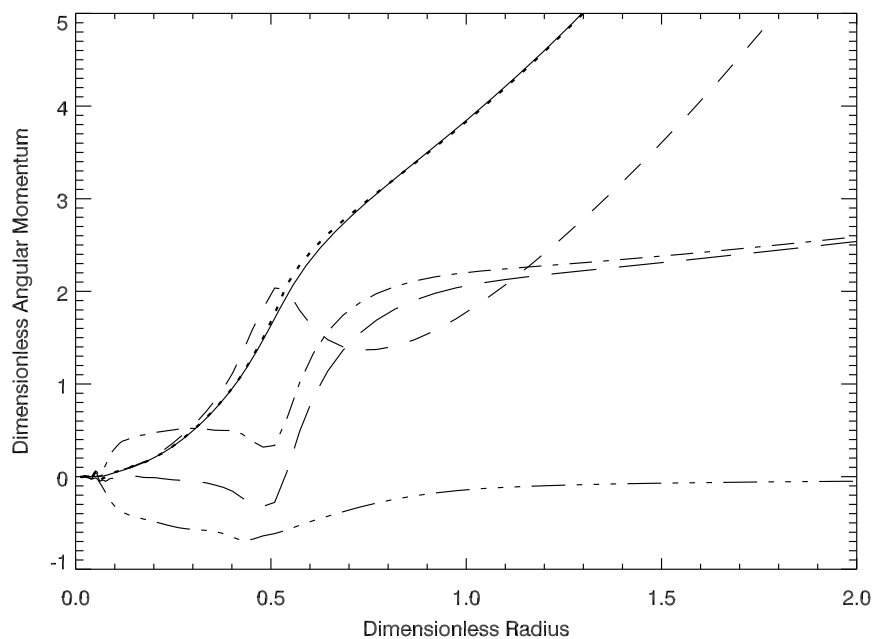}
\caption{Time averaged distributions of the dimensionless angular 
momentum $l(x)$ enclosed within an expanding sphere of fixed dimensionless 
radius $x$, computed in two different ways, using 
eqs.~(\ref{angmom_dimensionless}) (solid line) and
(\ref{angmom_cons_dimensionless}) (dotted). The averaging is done 
between $3.0\times 10^{11}$ and $7.5\times 10^{11}$~sec. Also 
plotted are the three terms on the right hand side of 
eq.~(\ref{angmom_cons_dimensionless}), showing the angular 
momentum change due to volume expansion (dashed), mass crossing the
sphere (dot-dashed), and magnetic braking (dot-dot-dashed). The 
sum of the last two terms is plotted as the long dashed line, 
showing that the angular momentum advected into a sphere of
radius $x~\lesssim ~0.4$ by fluid motion is almost completely removed 
by magnetic braking. 
}
\label{L_av}
\end{figure}

\begin{figure}
\epsscale{0.75}
\plotone{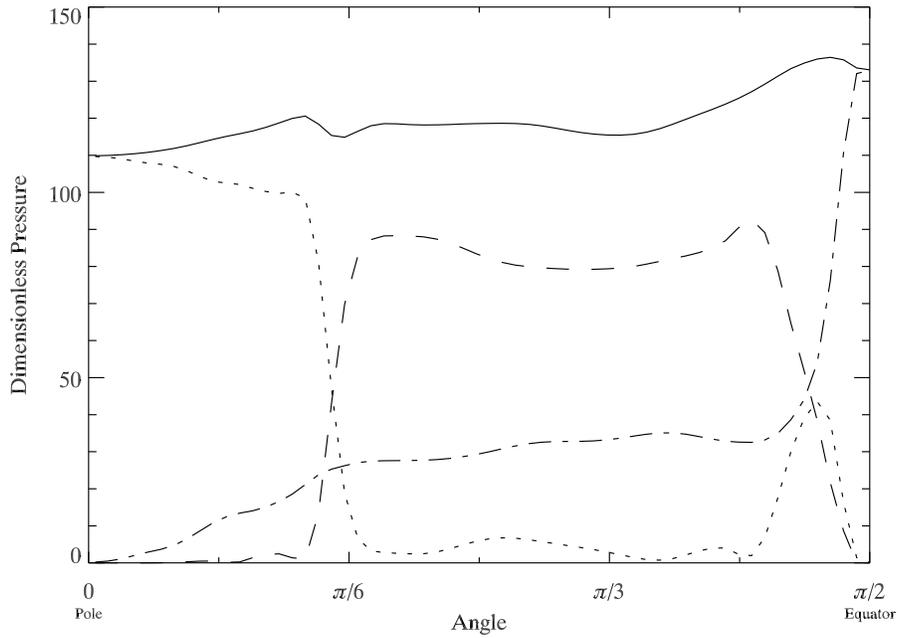}
\caption{Time averaged angular distributions of the total pressure and its
components at a representative radius $x=0.3$ for the standard
case. The total pressure (solid line) is nearly 
constant in $\theta$, but is dominated by different components in different 
regions. The thermal pressure (dot-dashed) dominates in the
pseudodisk, the toroidal 
magnetic pressure (dashed) dominates in the bubble, and the poloidal magnetic 
pressure (dotted) dominates in the polar region.  }
\label{pressures}
\end{figure}

\begin{figure}
\epsscale{0.75}
\plotone{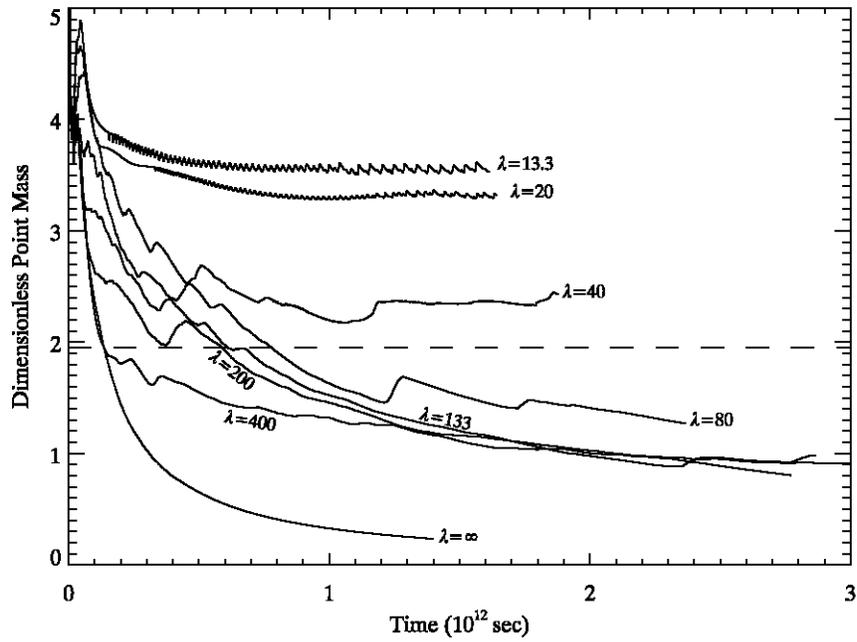}
\caption{Time evolution of dimensionless point mass for relatively weak 
magnetic field cases. The mass-to-flux ratio increases from the standard 
model ($\lambda=13.3$) on the top to the non-magnetized case ($\lambda=\infty$) 
on the bottom.}
\label{Mstar_LowB}
\end{figure}

\begin{figure}
\epsscale{1.0}
\plotone{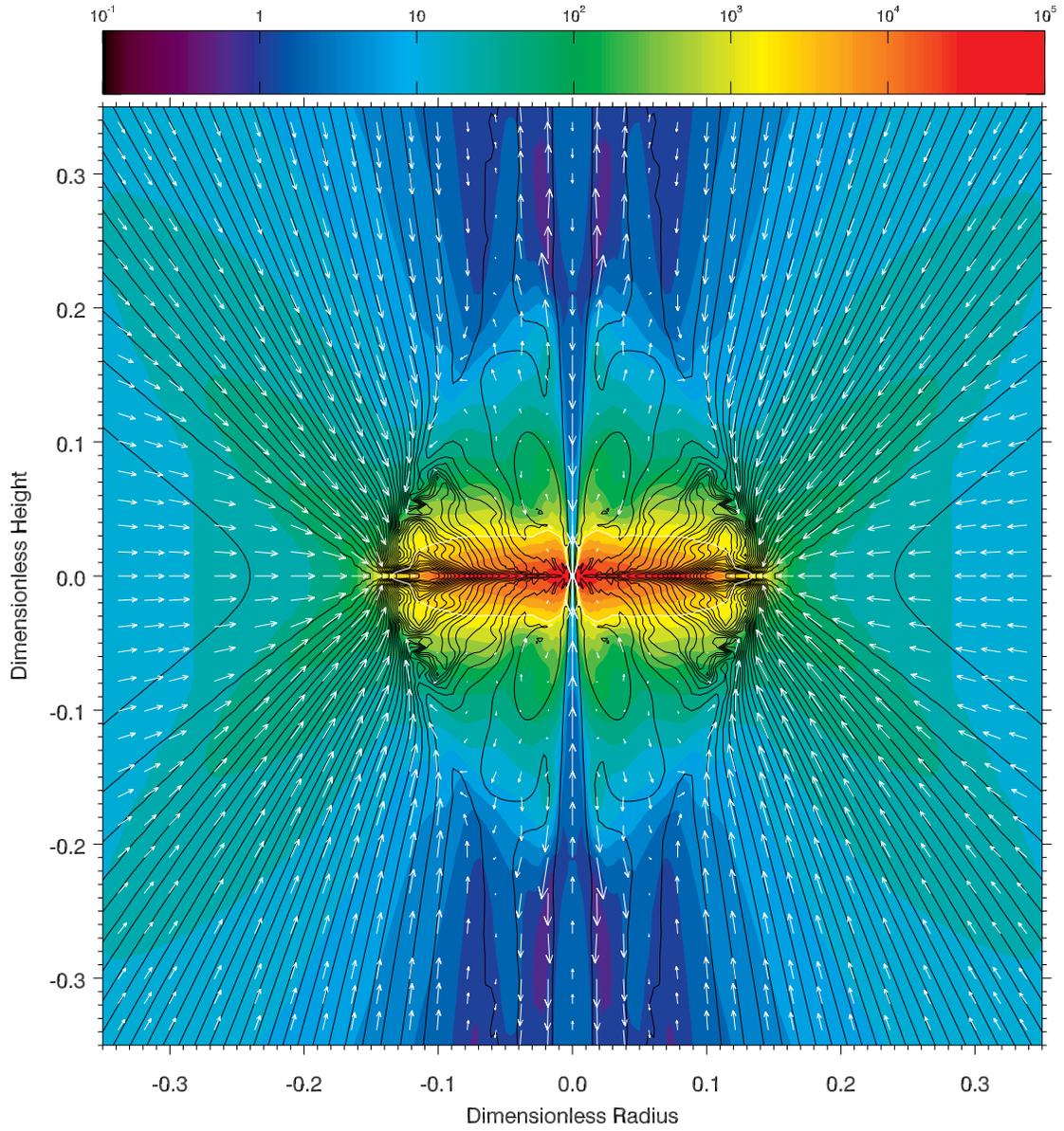}
\caption{A snapshot of the disk formed in the weakest field case 
of $\lambda=400$. The density is plotted in color contours, with 
the magnetic fields lines in black 
and the velocity field as white arrows. The dimensionless 
density of $3\times10^{3}$ (white contour) provides a fiducial  
boundary for the disk.}
\label{snap_Bpt01}
\end{figure}

\begin{figure}
\epsscale{0.75}
\plotone{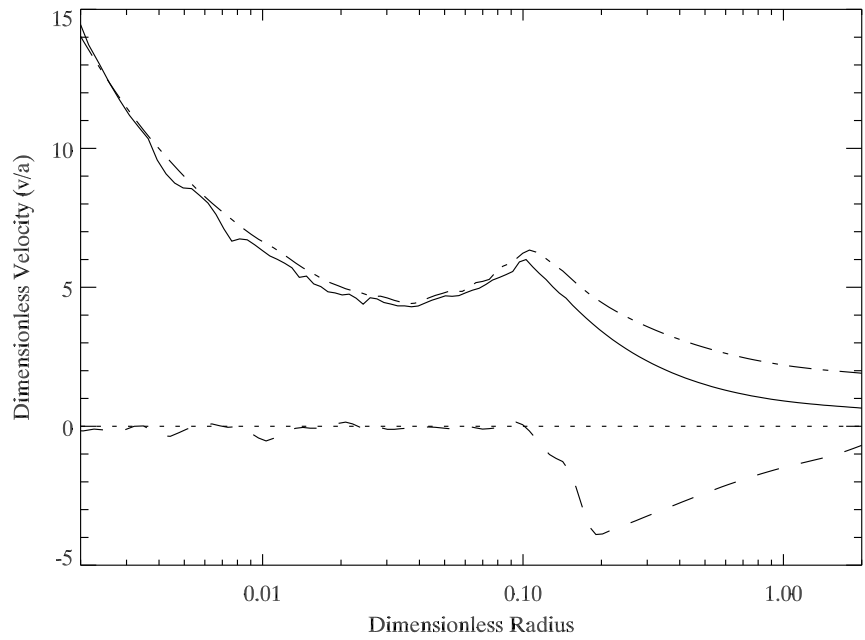}
\caption{Infall (dashed line) and rotation (solid)  speeds on the equator at the representative 
time for the weakest field case of $\lambda=400$. The rotational support for the disk is 
seen by comparing the rotation speed to the equilibrium value (dot-dashed), and by the 
corresponding drop in the infall speed. }
\label{Disk_speeds}
\end{figure}

\begin{figure}
\epsscale{0.75}
\plotone{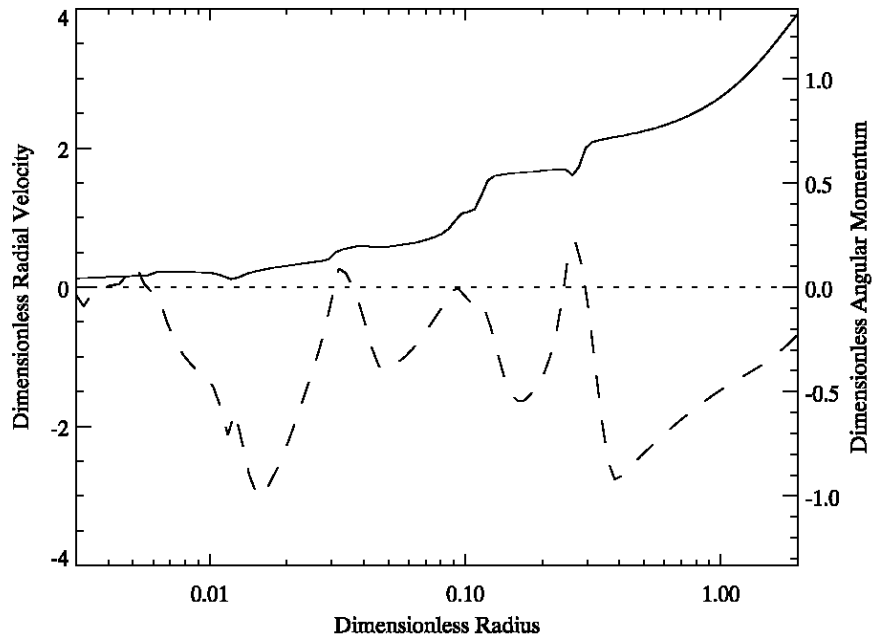}
\caption{Radial velocity (dashed curve) and specific angular momentum 
(solid) on the equator at a representative time $t=2\times
  10^{12}$~sec for the relatively weak field case of $\lambda=80$. 
They are in units of, respectively, the sound speed $a$ and $a^2 t$.  
Note the abrupt drops in angular momentum at the stagnation points of the
collapsing flow.}
\label{vr_specmom_Bpt05}
\end{figure}

\begin{figure} 
\epsscale{1.0}
\plotone{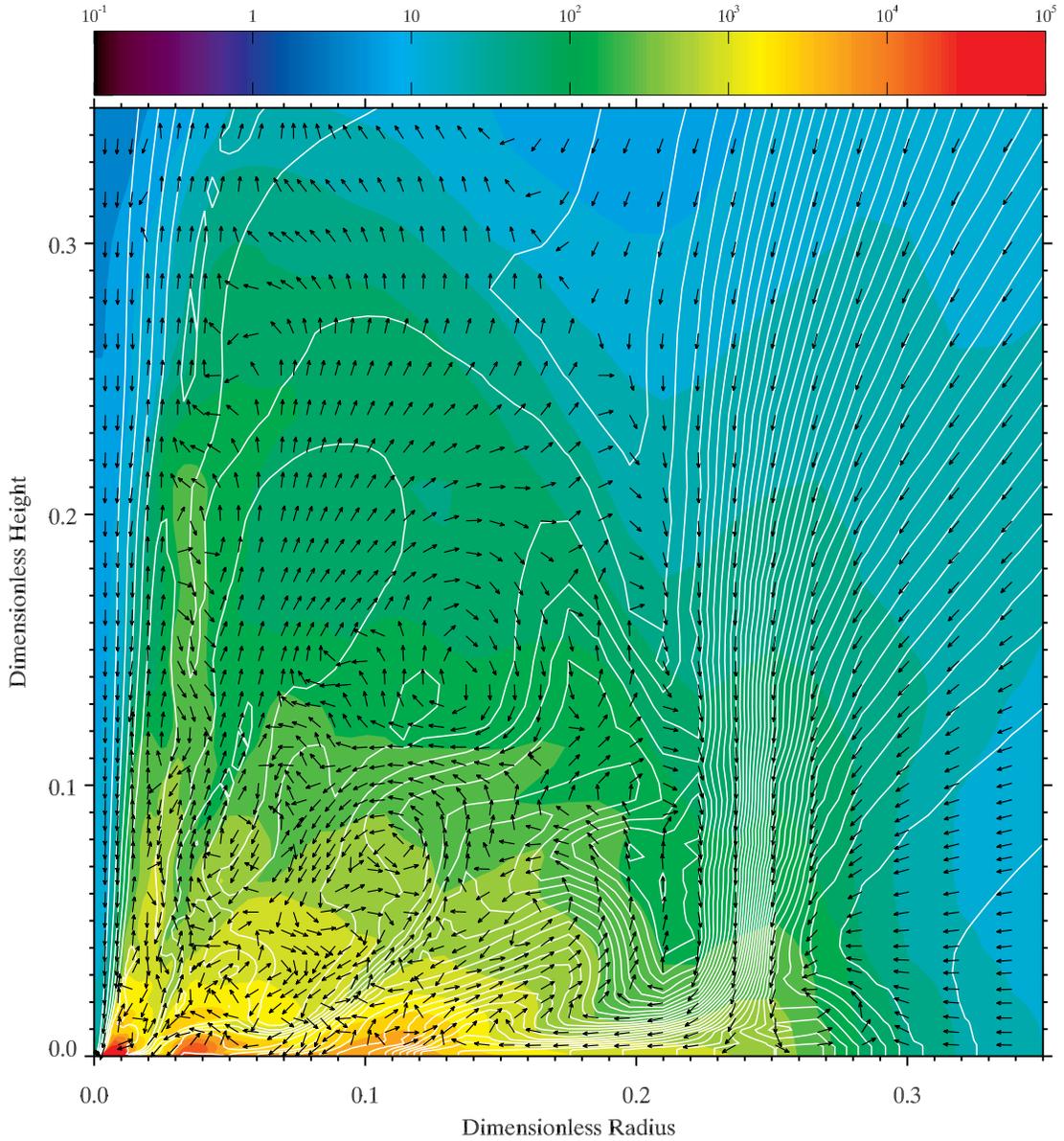}
\caption{Map of density distribution in a transition case ($\lambda=80$), 
at the representative time. Transient high density rings form instead of 
a disk as infalling material piles up near centrifugal radii, then
recollapses after losing a large fraction of its angular momentum 
through magnetic braking. White contours are magnetic field lines, 
and arrows are unit velocity vectors.}
\label{disk_unitvec_Bpt05}
\end{figure}

\begin{figure}
\epsscale{0.75}
\plotone{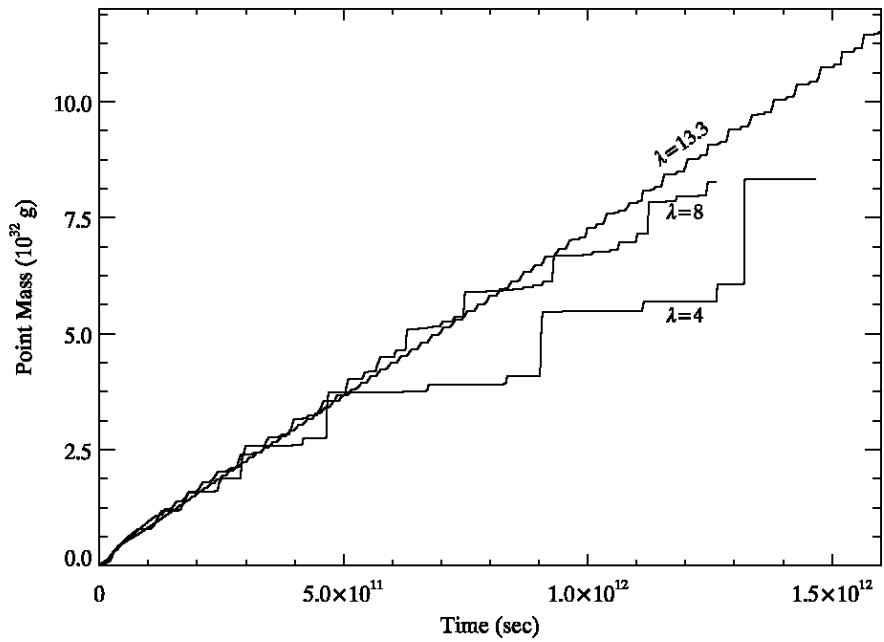}
\caption{Time evolution of dimensionless point mass for relatively strong
magnetic field cases. The mass-to-flux ratio decreases from top to
bottom, the opposite of the trend seen in the low field cases.  Note 
the step-like behavior, especially in the highest field case,
indicative of episodic mass accretion.}
\label{Mstar_HighB_noscaled}
\end{figure}

\begin{figure}
\epsscale{1.0}
\plotone{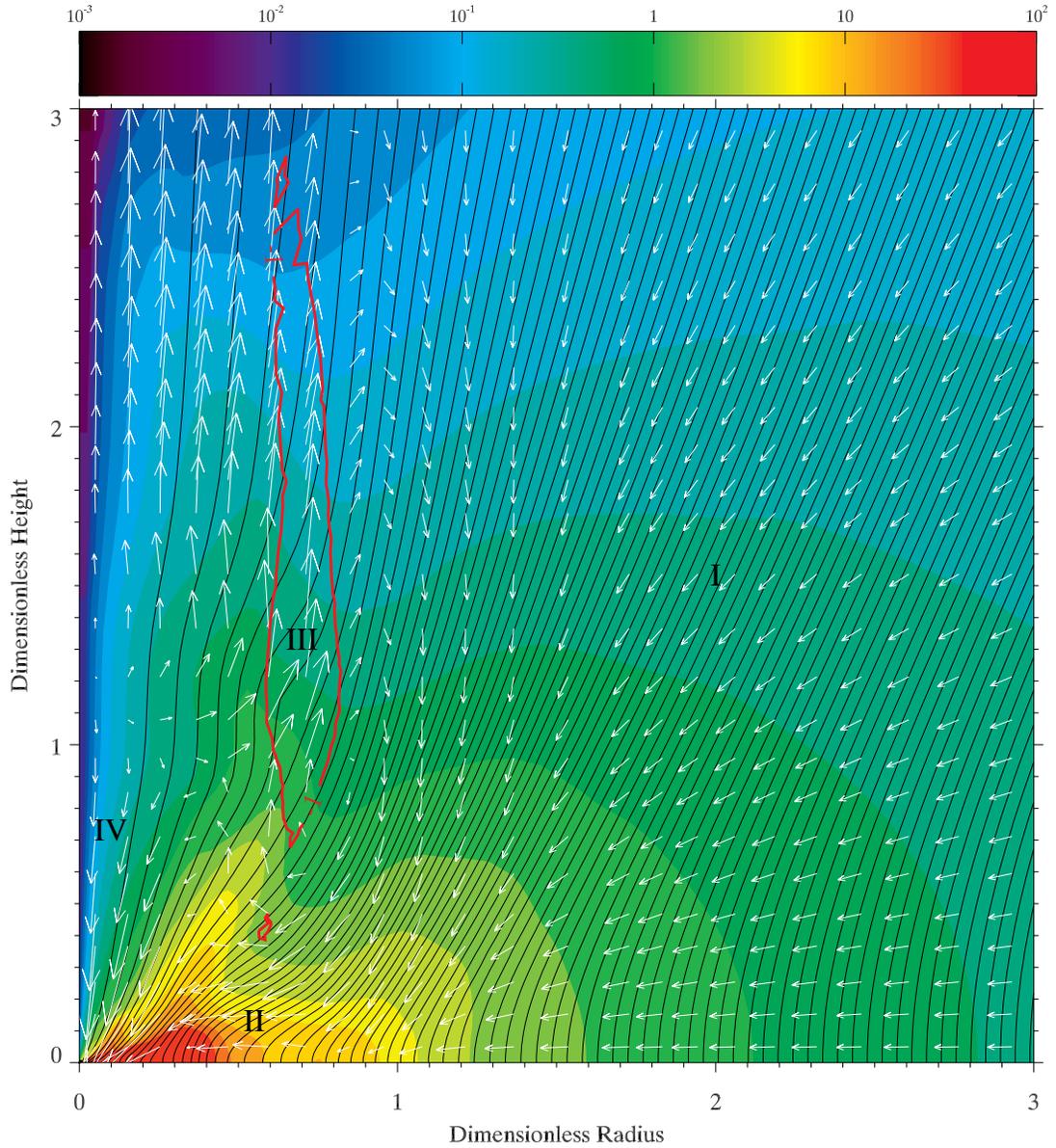}
\caption{Snapshot of the strongly magnetized ($\lambda=4$) core 
representative of the state before a ``reconnection'' event. It is to 
be compared with the standard case shown in Fig.~\ref{snap}. 
Enclosed within the red contour is the region where the magnetic
field is dominated by the toroidal component. The four dynamically 
distinct regions are labelled here as in Fig.~\ref{snap}.}
\label{snap_B1pt0}
\end{figure}

\begin{figure}
\epsscale{1.0}
\plotone{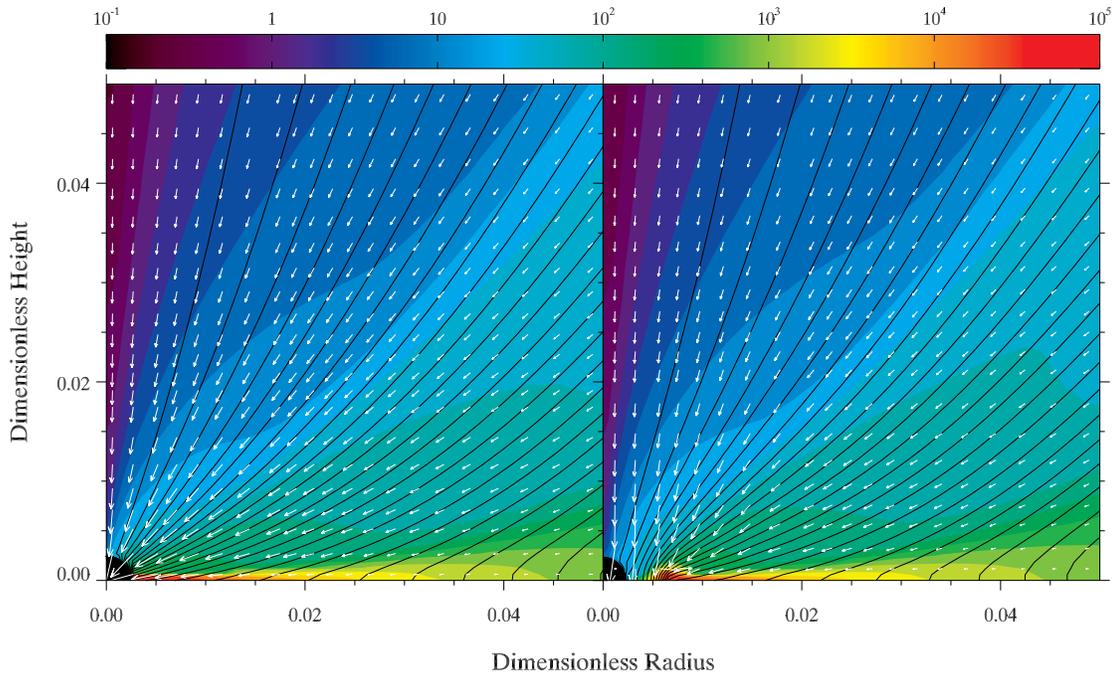}
\caption{Density distribution (color map), velocity field (white arrows) and
  magnetic field lines (black lines) before (left panel) and after
  (right) a reconnection event. 
}
\label{burst}
\end{figure}

\begin{figure}
\epsscale{0.75}
\plotone{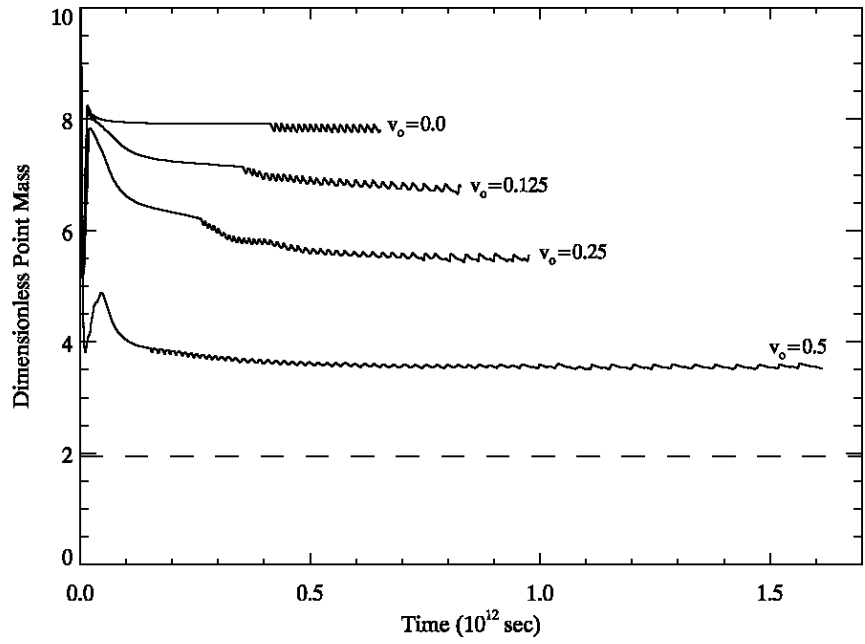}
\caption{Dimensionless central point masses for collapsing cores of the 
same initial magnetic field ($\lambda=13.3$) but different rotation 
speeds. The accretion rate decreases as the rotation rate increases.  
Also plotted for comparison is the point mass for SIS (dashed line).}
\label{Mstar_diffrot}
\end{figure}

\begin{figure} 
\epsscale{1.0}
\plotone{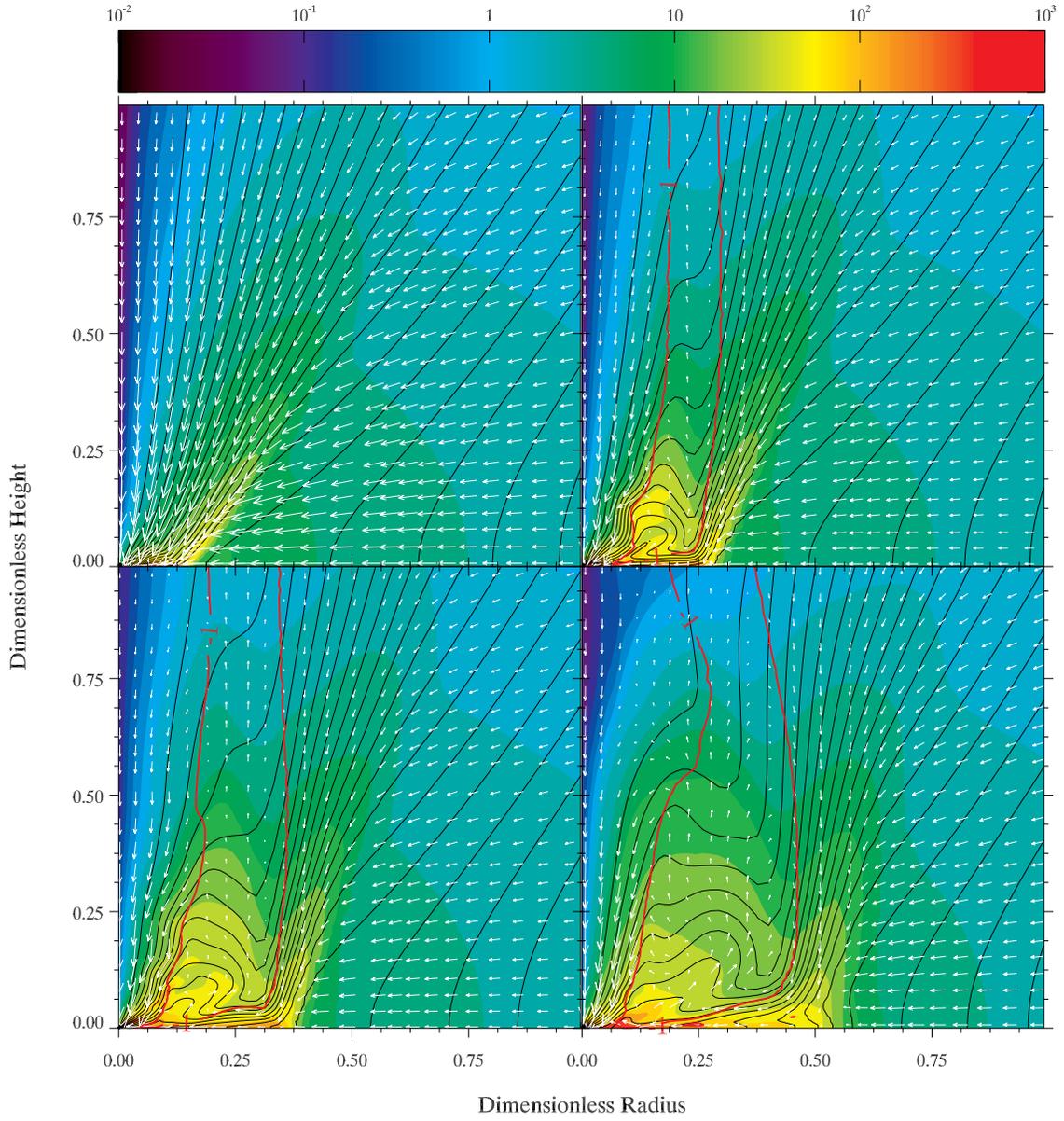}
\caption{Snapshots of collapsing cores of the same initial magnetic field but
different rotation speeds of $v_o=0$ (upper left panel), 0.125 (upper
right), 0.25 (lower left), 0.5 (lower right). The magnetic bubble,
enclosed  within the red contour, becomes thinner with less rotation, 
disappearing entirely in the non-rotating case.}
\label{multi_diffrot}
\end{figure}

\begin{figure}
\epsscale{1.0}
\plotone{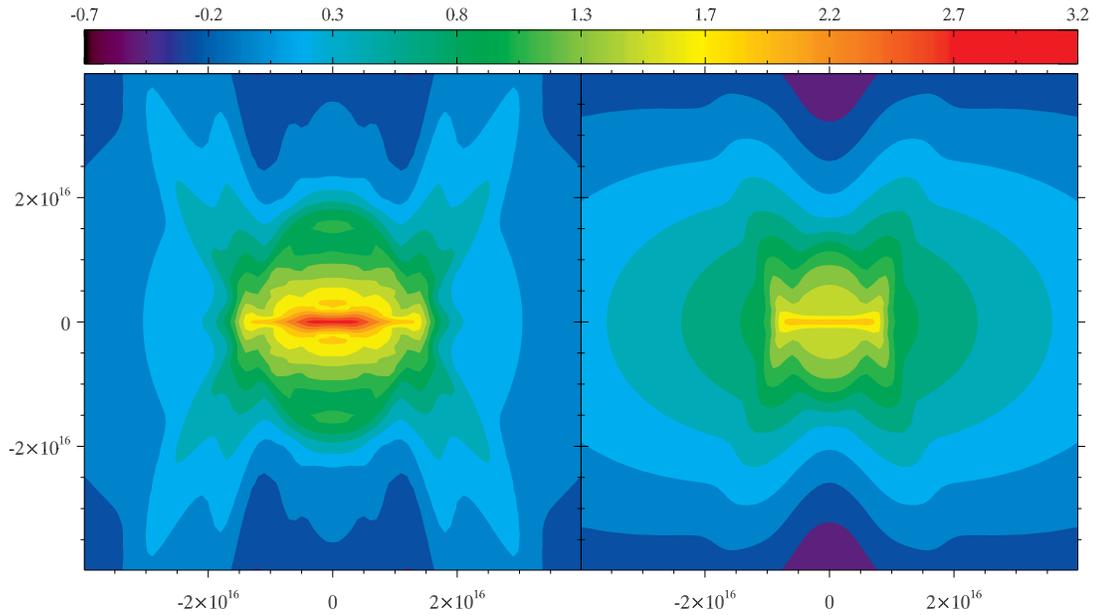}
\caption{Column density distributions for the $\lambda=80$ case 
at $t=2\times 10^{12}$~sec (left panel) and the $\lambda=13.3$ 
case at $t=8.3\times 10^{11}$~sec (right) case viewed perpendicular 
to the axis. In each case, a dense equatorial pseudodisk is 
surrounded by an extended structure supported by a combination 
of magnetic fields and rotation --- a magnetogyrosphere. 
The length and column density are in units of $cm$ and 
$g~cm^{-2}$ respectively.  
} 
\label{columndensity}
\end{figure}

The infalling envelope can also be stopped at the magnetic barrier. 
The barrier is a general feature produced by the poloidal magnetic 
field lines draping over the outer surface of the magnetogyrosphere 
(see Fig.~\ref{snap} and 
Fig.~\ref{disk_unitvec_Bpt05}). The field lines channel the envelope 
material towards the equator, where it is slowed down temporarily, 
before recollapsing towards the center. In the case of IRAM 04191, 
there is evidence that the material interior to the stagnation radius 
$r_0$ is moving inwards at a speed comparable to the sound speed 
\citep{2005ApJ...619..948L}. The rapid infall implies that there is not enough 
angular momentum to hold up the material by rotation. Indeed, there 
is some hint that the region interior to $r_0$ rotates more or less 
as a solid body, which would point to a strong braking, as expected
in a magnetic barrier, where the field is strong and the slowdown 
of infall allows more time for the magnetic braking to operate. 
The braking is apparently so efficient that an 
optically thick disk does not form until within 10~AU of the central 
object, if at all \citep{2002A&A...393..927B}. 

If strong magnetic braking is indeed operating inside the stagnation
radius $r_0$ of IRAM 04191, where is the bulk of the extracted angular 
momentum deposited? 
One possibility is the slowly outflowing region (with radial speeds 
of order 0.4~km/s) that shows up in the position-velocity diagram 
along the minor axis of the ring-like $N_2H^+$ structure \citep{2005ApJ...619..948L}. 
This 
region could be part of the magnetogyrosphere. Alternatively, it 
could be material accelerated outwards by the fast protostellar 
wind \citep{2005ApJ...619..948L}. The ambiguity illustrates one of the 
observational challenges in probing the magnetogyrosphere directly;
it may easily be masked by the wind-driven motions. This problem 
may be alleviated using molecules that are little affected by 
wind-interaction. Another complication in direct probing of the 
magnetogyrosphere is the depletion of molecules onto dust grains. 
Even though $N_2H^+$ is among the last molecules to disappear at 
high densities \citep{2002ApJ...569..815T}, it is heavily 
depleted close to the central object in 
IRAM 04191 \citep{2004A&A...419L..35B,2005ApJ...619..948L}. 

In the absence of detailed kinematic information, the morphology 
of circumstellar material may provide indirect evidence for the 
magnetogyrosphere. A possible example is the HH 211 system, the 
prototype of the class of highly collimated CO molecular outflows 
\citep{1999A&A...343..571G,1999osps.conf..227B}. It has a pair of 
oppositely directing, fast-moving jets, each enclosed by a narrow, 
slower shell. Recent PdBI observations of the source at a 
resolution of $0.35^{\prime\prime}$ reveals that the outer shell 
is much narrower within about $10^3$~AU of the central object 
(indeed unresolved transversely, Gueth et al., in preparation). 
It appears that, as the outflow propagates away from the central 
region, it is initially confined laterally by a $10^3$~AU-scale 
structure before suddenly expanding into the more widely open 
shell-like structure observed at larger distances. A natural 
possibility for the structure is the magnetogyrosphere, which 
can provide confinement through both thermal pressure and magnetic 
``hoop'' stresses associated with the toroidal field. Indeed, a 
glance at the density distribution in Fig.~\ref{disk_unitvec_Bpt05} 
(or Fig.~\ref{snap}) reveals a narrow funnel in the polar region 
close to the origin where the outflow can plausibly be confined. 
To strengthen the case for magnetogyrosphere as the confining 
structure at the base of the HH 211 molecular outflow, high resolution 
interferometric observations that probe the kinematics (and ideally 
the magnetic structure) of the confining structure are needed. 

To summarize, there is tantalizing evidence for a magnetic braking-driven 
magnetogyrosphere in IRAM 04191 and, to a lesser extent, HH 211. To 
make a stronger case, more high resolution kinematic data are needed, 
perhaps from the recently completed CARMA and upgraded PdBI, and 
especially the upcoming ALMA. A unique model prediction is the 
existence of more than one centrifugal barriers interior to the 
magnetic barrier, depending on the field strength and assuming 
good coupling between the field and matter. Detection of such a 
feature would clinch the case for disk disruption by (repeated) 
magnetic braking. 

\subsubsection{Dispersal of Magnetogyrosphere and Late-Time Disk Formation} 
\label{dispersal}

If the bulk of the angular momentum of the collapsed material is 
indeed deposited in the magnetogyrosphere, what eventually happens 
to the magnetogyrosphere becomes an important issue in the 
angular momentum evolution of the system. We believe the fate of the 
magnetogyrosphere is tied to that of the dense core material, 
the majority of which is not incorporated into stars. For example, 
in the Taurus molecular clouds, the dense HCO$^+$ cores typically contain 
a few to several solar masses \citep{2002ApJ...575..950O}, whereas 
the typical mass of 
the YSOs is only $\sim 0.5$~M$_\odot$ \citep{1995ApJS..101..117K}. It is
likely that the bulk of the core material is removed by the 
powerful protostellar wind \citep{1987ARA&A..25...23S,2000ApJ...545..364M}. 
Since the specific angular momentum of the core material 
tends to increase with distance, it follows that the protostellar 
wind blows away the majority of not only the mass, but also the 
angular momentum, of the system, independent of magnetic braking 
and magnetogyrosphere formation. 

The formation of a magnetogyrosphere in place of a rotationally
supported disk can make the wind-stripping of angular momentum 
potentially more efficient; it should be easier to blow away the 
more spherical, lower density, magnetogyrosphere than a disk. 
The dispersal 
of the magnetogyrospheric material allows the wind to remove the 
angular momentum associated with not only the portion of the 
core material that does not collapse into the central object but 
also the collapsed portion; the latter is parked mostly in the 
magnetogyrosphere. Wind-stripping is a powerful mechanism of 
angular momentum removal that deserves close attention. 

The total angular momentum retained in a stellar system may be 
determined 
to a large extent by the amount of the high specific angular 
momentum magnetospheric and/or core material that survives the 
wind-stripping. The left-over material may recollapse towards 
the center, perhaps forming a rotationally supported disk, which 
could persist once the massive, slowly rotating envelope that  
strongly brakes the disk rotation at earlier times has been 
cleared away. It takes only a small fraction of the mass and 
angular momentum of a typical dense core to form a 
minimum solar nebula. Removal of the braking material --- the 
slowly rotating envelope --- is one way for the all-important, 
rotationally supported disk to appear in the collapse of a 
magnetized core. Another possibility, to 
be discussed briefly in \S~\ref{future} below, is through nonideal
MHD effects, such as ambipolar diffusion.

\subsection{Low Plasma-$\beta$ Disks Supported by Rotation} 
\label{lowbetadisk}

\begin{figure}
\epsscale{0.75}
\plotone{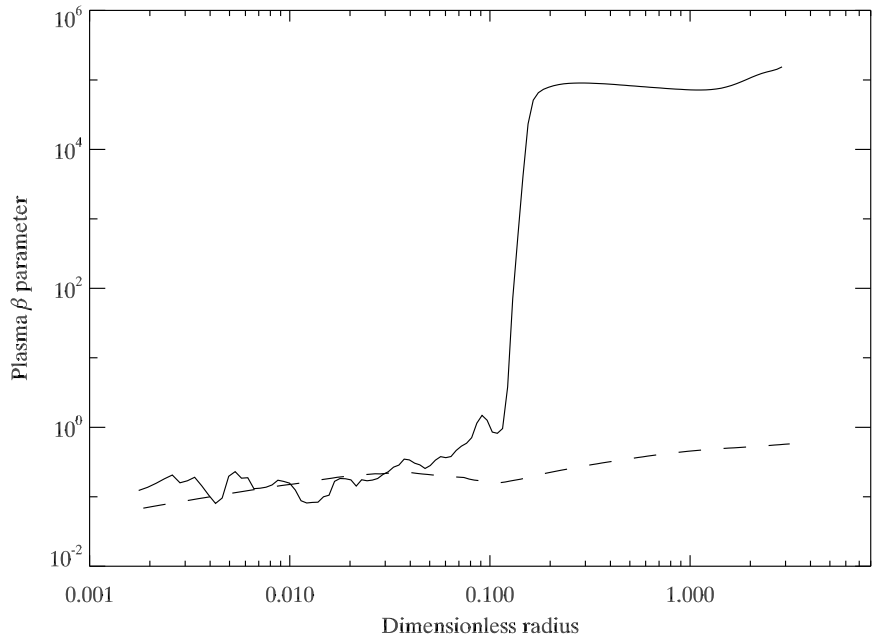}
\caption{Plasma-$\beta$ parameter along a direction $3^\circ$ away from 
the midplane at a representative time $t=2\times 10^{12}$~sec (solid 
line). Plotted for comparison is a plausible limiting value estimated 
analytically (dashed line). See text for discussion.} 
\label{beta}
\end{figure}

We find persistent, rotationally-supported, equilibrium disks 
only in the three 
weakest field cases, with $\lambda=400$, $200$ and $133$. These 
mass-to-flux ratios correspond to initial values of equatorial 
plasma-$\beta$ of $7.68\times 10^4$, $1.92 \times 10^4$ and $8.53 
\times 10^3$, respectively. One may not expect such weak fields 
to have any significant effect on the dynamics. However, they 
are strongly amplified, through both infalling motion and 
differential rotation. The amplification is so efficient that, 
even in the weakest field case of $\lambda=400$, the magnetic 
pressure dominates the thermal pressure everywhere in the disk 
(except near the equatorial plane, where the 
toroidal field is forced to zero by symmetry), especially in the 
inner part, where the orbital period is shorter. The enormous 
difference in the plasma-$\beta$ between the disk region and the 
collapsing envelope is illustrated in Fig.~\ref{beta}, where we 
plot $\beta$ as a function of radius along a direction $3^\circ$  
away from the midplane. The distribution is similar along other 
angles, including those far from the midplane, where the amplified 
field creates a vertically extended, magnetic tower. The base of 
the tower is a magnetically dominated ``corona'' that replaces 
the classical ``accretion shock'' that bounds the disk in the 
absence of magnetic fields \citep[e.g.,][]{1990ApJ...355..651B}. It 
can be viewed as a weak version of the magnetogyrosphere; the 
bulk of the mass and angular momentum of the collapsed material 
reside in the disk rather than the corona.  

The disk magnetic field cannot be amplified indefinitely. 
\citet{2007MNRAS.375.1070B} 
conjectured that the limiting field strength is 
such that the Alfv\'en speed $V_A$ roughly equals the geometric 
mean of the Keplerian speed $V_K$ and the sound speed $a$ of the 
gas. The conjecture is motivated by the 
analysis of \citet{2005ApJ...628..879P}, who showed that the growth 
rate of the magnetorotational instability vanishes when 
$V_A=(2 V_K a)^{1/2}$. It is qualitatively consistent with our 
finding that the plasma-$\beta$ tends to be lower in the inner 
part of the disk where the rotation speed is higher. For a 
quantitative comparison, we plot in Fig.~\ref{beta} the predicted 
plasma $\beta$ distribution based on 
$\beta= 2 a^2/V_A^2=a/V_K$, where $V_K$ is the equilibrium rotation 
speed shown in Fig.~\ref{Disk_speeds}. The rough agreement between 
the predicted and actual distributions lends some support to the 
conjecture, even though the toroidal field in our simulation is 
generated mostly by simple twisting of a pre-existing poloidal 
field, rather than MRI. 

The condition for a rotationally supported disk {\it not} to be 
affected significantly by magnetic braking in the 
ideal MHD limit, $\lambda \gtrsim 100$, may be difficult to satisfy 
in the Galaxy at the present time. For a typical low-mass core, 
the field strength must be less than $\sim 1~\mu$G on the 0.05pc
scale. The required strength is 
well below the median value inferred by
\citet{2005ApJ...624..773H} for the cold neutral structures of HI gas. 
A potential exception is the massive cores of molecular clouds, 
where column densities as high as 1~g~cm$^{-2}$ are often inferred 
\citep{1997ApJ...476..730P}. If their field strength is about $15$~$\mu$G 
or less, the mass-to-flux ratio $\lambda$ would be of order 100 
or more. Such a weak field (relative to mass) should not significantly 
disrupt disk formation, particularly around 
massive stars that may form in such cores \citep{2002Natur.416...59M}, 
although it can still be amplified to the extent that its pressure 
dominates the thermal pressure inside the disk, at least in 2D. 
If the cores are more strongly
magnetized, as is the case for the nearest region of ongoing 
massive star formation OMC1 \citep{1999ApJ...520..706C,2004ApJ...616L.111H}, 
the field may suppress the formation of a rotationally supported
disk altogether, as long as it remains well coupled to the 
matter. In this case, most of the mass would be accreted through 
a still rotating but rapidly infalling pseudodisk that is 
strongly braked by magnetic fields. Quantifying this possibility 
for massive star formation fully would require 3D simulations 
that include both radiative transfer \citep*[e.g.,][]{2007ApJ...656..959K} 
and magnetic fields, as well as a treatment of ionization 
and magnetic coupling.

\subsection{Future Directions} 
\label{future}

We have performed calculations of disk formation in rotating dense 
cores magnetized to different degrees under a number of simplifying
assumptions. These include axisymmetry, ideal MHD, and ignoring 
protostellar winds. The calculations provide a starting point 
for future refinements. 

The most straightforward refinement is to increase spatial resolution. 
Higher resolution 2D (axisymmetric) simulations are desirable in the 
weakest field cases, where rotationally-supported, equilibrium 
disks are formed. 
Such disks can evolve through the twisting of a pre-existing, poloidal 
field dragged into the disk from large distances by collapsing
material. The disk evolution can be aided, perhaps even dominated, by 
MRI. To adequately quantify the role of MRI in driving disk evolution, 
one need to resolve the fastest growing MRI mode
\citep{2001ApJ...548..348H,2001MNRAS.322..461S}, whose wavelength 
decreases as the 
field weakens. Another straightforward refinement is to relax the mirror 
symmetry imposed at the equator. We have explored a few cases without 
the equatorial symmetry, and found qualitatively similar results. 
They will be discussed in the future along with 3D simulations. 

High resolution, global 3D simulations that include both the core 
and the disk will be a natural, but challenging, extension of our 
weak field calculations. They are needed for determining the 
importance of gravitational torques in redistributing the disk 
angular momentum relative to hydromagnetic stresses \citep[e.g.,][]{2004ApJ...616..364F}, 
including those arising from possible MRI associated 
with the dominant, toroidal component of the disk magnetic field 
that cannot be treated in 2D. 3D calculations are also needed for 
treating potential (sub-)stellar companion formation through 
fragmentation \citep[e.g.,][]{2006ApJ...641..949B,2007ApJ_MIM_sub,2007A&A_HT_sub}.

Equally challenging will be detailed treatment of nonideal MHD 
effects. Such effects have been investigated over the years, 
most systematically by Nakano and collaborators 
\citep*[e.g.,][]{2002ApJ...573..199N}, but have yet to be incorporated 
into multi-dimensional {\it protostellar} collapse calculations. 
Particularly vulnerable to non-ideal effects are two prominent
features of our relatively 
strong field cases: the split magnetic monopole near the center 
and the associated episodic mass accretion. Sharp kinks in field 
lines (across the midplane), such as those shown in the left 
panel of  Fig.~\ref{burst}, are expected to be smoothed out 
by nonideal MHD effects \citep[see][for an example]{2006ApJ...647..382S}. 
However, there will still be a 
tendency for the magnetic flux to accumulate at small radii even 
when realistic nonideal effects are present. Indeed, episodic 
mass accretion due to repeated flux 
accumulation and escape can be seen in the nonideal MHD calculations of 
\citet{2005ApJ...618..783T} under the thin-disk 
approximation.

Whether the deviation from ideal MHD in a collapsing core is strong 
enough to enable disk formation is uncertain. \citet{2002ApJ...580..987K} 
investigated the effects of ambipolar diffusion on magnetic
braking and disk formation under the self-similar assumption and 
thin-disk approximation. They showed that rotationally supported 
disks can indeed form even in strongly magnetized cores, although
this is by no means guaranteed. A particular complication is the 
hydromagnetic accretion shock of C-type produced by the magnetic 
flux left behind by the mass that has gone into the central object 
\citep{1996ApJ...464..373L,1998ApJ...504..257C}. The increase in field
strength and decrease in infall speed inside the shock and in the
postshock region tend to enhance the magnetic braking 
\citep[in a manner akin to the quasistatic phase of subcritical cloud evolution,][]{1994ApJ...432..720B}, 
whereas the slippage of field lines relative 
to neutral matter in the azimuthal direction tends to decrease 
it. Whether there is enough angular momentum left in the accreting 
material after passing through the C-shock to form an appreciable 
disk is uncertain, as demonstrated by \citet{2002ApJ...580..987K} 
semi-analytically in 1D. Numerical study of this problem in 2D is 
underway (Mellon \& Li, in preparation). 

Theory of disk formation will be incomplete without a treatment of 
protostellar wind. It is likely that the wind removes the bulk of 
not only the mass, but also the angular momentum, of a (low-mass) 
star forming core, at least for regions like the Taurus clouds, where
the typical mass of a core is well above that of a YSO. As 
discussed in \ref{dispersal}, there is the possibility of late-time 
disk formation after the wind has stripped away most of the (slowly 
rotating) envelope that impedes disk formation through magnetic 
braking. The role of protostellar wind in the evolution of angular
momentum in star formation in general and disk formation in particular
remains to be quantified. 
The various refinements outlined above are not mutually 
exclusive. However, it will be a daunting, if not impossible, task to 
implement all of them at the same time. Fortunately the calculations 
can be guided by high resolution observations with increasingly 
powerful interferometers, especially ALMA. 

\section{Conclusions}
\label{conclusion}

We have performed a set of axisymmetric calculations of protostellar
collapse phase of star formation in rotating cores magnetized to
different degrees, expanding on the work of ALS03. The goal is to 
determine how magnetic braking affects disk formation in the ideal 
MHD limit. The main conclusions are: 

1. Protostellar disks can be disrupted by weak magnetic fields 
corresponding to mass-to-flux ratio $\lambda \gg 1$. This is made 
possible by the tight wrapping of the disk field lines, which increases 
the rate of magnetic braking, and the long braking time compared with 
the disk rotation period. In our illustrative calculations, the disk 
disruption begins around $\lambda~ \sim 100$.

2. Magnetic braking disrupts the rotationally supported disk by
creating regions of rapid, supersonic collapse in the disk. These
regions are separated by one or more centrifugal barriers, where 
the rapid infall is halt temporarily and the magnetic braking 
is locally enhanced. The number of centrifugal barriers decreases 
as the core becomes more strongly magnetized. Multiple centrifugal 
barriers, if observed, would clinch the case for disk disruption 
by magnetic braking.

3. Surrounding the strongly braked disk of spatially alternating 
collapse and stagnation is a vertically extended 
structure supported by a combination of (toroidal) magnetic field 
and rotation --- a ``magnetogyrosphere'' --- where the bulk of the 
angular momentum of the collapsed material is stored. The 
magnetogyrosphere may be detectable interferometrically, although 
the detection may be complicated by interaction with protostellar 
winds. We suggest that the wind may play a key role in determining 
not only the mass but also the angular momentum of a stellar system.

4. Even in the extremely weak field cases of $\lambda~ \gtrsim~ 100$,
where a rotationally-supported, equilibrium disk is formed in 
the ideal 
MHD limit, the magnetic field can still be dynamically important. 
It can be amplified to the extent that its pressure dominates the 
thermal pressure both in the disk and its surrounding region, at 
least in 2D. The field may affect the fragmentation properties 
of the disk, and seed the MRI, although high resolution 3D 
simulations are required to fully quantify its dynamical effects. 

5. In relatively strongly magnetized cores of $\lambda~ \lesssim~ 10$, the
disk formation is completely suppressed, and the bulk of the angular
momentum of the collapsed material is ejected out of the system 
via a low-speed, magnetic braking-driven wind, as found previously. 
A new feature is that the mass accretion on to the central object 
is highly episodic, because of magnetic reconnection. How this and 
other behaviors are modified by nonideal MHD effects, particularly
ambipolar diffusion, remains to be quantified. 

In summary, our calculations have shown that in the ideal MHD limit 
the magnetic braking is strong enough to prevent the formation 
of a persistent, rotationally supported, protostellar disk in all 
but the unrealistically weak field cases. Since disks are observed 
around many, if not all, young stellar objects, this result presents 
an interesting conundrum that must be resolved. For such a disk to 
form in dense cores magnetized to a realistic level, magnetic 
braking must be weakened one way or another, perhaps through a 
combination of nonideal MHD effects and protostellar winds. 
Quantifying these effects should be a major goal of future research 
in this area. 

\acknowledgements
This work was supported in part by NASA (NNG05GJ49G) and NSF (AST-0307368)
grants. It grew out of an earlier collaboration that involved Tony 
Allen and Frank Shu. We thank Tony Allen for technical assistance at 
the beginning of the project, Frank Shu, John Hawley, and Philippe
Andre for helpful 
discussion, and Sebastien Fromang for providing the gravity solver 
used in our work.

\clearpage

\begin{thebibliography}{}
\bibitem[Allen et al.(2003)Allen, Li, \& Shu]{2003ApJ...599..363A} Allen, A., Li, Z.-Y., \& Shu, F.~H.\ 2003, \apj, 599, 363 
\bibitem[Andr{\'e} et al.(1999)Andr{\'e}, Motte, \& Bacmann]{1999ApJ...513L..57A} Andr{\'e}, P., Motte, \ F., \& Bacmann, A.\ 1999, \apjl, 513, L57 
\bibitem[Bachiller \& Tafalla(1999)]{1999osps.conf..227B} Bachiller, R., \& Tafalla, M.\ 1999, in NATO ASIC Proc.~540, The Origin of Stars and Planetary Systems, ed Lada, C.~J. and Kylafis, N.~D. (Norwell, MA: Kluwer Acad. Pub.), 227 
\bibitem[Balbus \& Hawley(1998)]{1998RvMP...70....1B} Balbus, S.~A., \& Hawley, J.~F.\ 1998, Reviews of Modern Physics, 70, 1 
\bibitem[Banerjee \& Pudritz(2006)]{2006ApJ...641..949B} Banerjee, R., \& Pudritz, R.~E.\ 2006, \apj, 641, 949 
\bibitem[Basu \& Mouschovias(1994)]{1994ApJ...432..720B} Basu, S., \& Mouschovias, T.~C.\ 1994, \apj, 432, 720 
\bibitem[Beckwith \& Sargent(1993)]{1993prpl.conf..521B} Beckwith, S.~V.~W., \& Sargent, A.~I.\ 1993, in Protostars and Planets III, ed. E.~H. Levy \& J. Lunine (Tucson:Univ of Arizona Press), 521 
\bibitem[Begelman \& Pringle(2007)]{2007MNRAS.375.1070B} Begelman, M.~C., \& Pringle, J.~E.\ 2007, \mnras, 375, 1070 
\bibitem[Belloche \& Andr{\'e}(2004)]{2004A&A...419L..35B} Belloche, A., \& Andr{\'e}, P.\ 2004, \aap, 419, L35 
\bibitem[Belloche et al.(2002)]{2002A&A...393..927B} Belloche, A., \ Andr{\'e}, P., Despois, D., \& Blinder, S.\ 2002, \aap, 393, 927 
\bibitem[Bodenheimer(1995)]{1995ARA&A..33..199B} Bodenheimer, P.\ 1995, \araa, 33, 199 
\bibitem[Bodenheimer et al.(1990)]{1990ApJ...355..651B} Bodenheimer, P., Yorke, H.~W., Rozyczka, M., \& Tohline, J.~E.\ 1990, \apj, 355, 651 
\bibitem[Boss(1998)]{1998ASPC..148..314B} Boss, A.~P.\ 1998, in ASP Conf. Ser. 148, Origins, ed. C.~E. Woodward, J.~M. Shull, \& H.~A. Thronson, Jr. (San Francisco: ASP), 314 
\bibitem[Ciolek \& K{\"o}nigl(1998)]{1998ApJ...504..257C} Ciolek, G.~E., \& K{\"o}nigl, A.\ 1998, \apj, 504, 257 
\bibitem[Cohl \& Tohline(1999)]{1999ApJ...527...86C} Cohl, H.~S., \& Tohline, J.~E.\ 1999, \apj, 527, 86 
\bibitem[Crapsi et al.(2007)]{2007A&A...470..221C} Crapsi, A., Caselli, P., Walmsley, M.~C., \& Tafalla, M.\ 2007, \aap, 470, 221 
\bibitem[Crutcher(1999)]{1999ApJ...520..706C} Crutcher, R.~M.\ 1999, \apj, 520, 706 
\bibitem[Crutcher \& Troland(2000)]{2000ApJ...537L.139C} Crutcher, R.~M., \& Troland, T.~H.\ 2000, \apjl, 537, L139 
\bibitem[Crutcher \& Troland(2007)]{2007IAUS..237..141C} Crutcher, R.~M., \& Troland, T.~H.\ 2007, in IAU Symp. 237, Triggered Star Formation in a Turbulent ISM, ed. B.~G. Elmegreen \& J. Palous (Cambridge: Cambridge Univ. Press), 141
\bibitem[Dunham et al.(2006)]{2006ApJ...651..945D} Dunham, M.~M., et al.\ 2006, \apj, 651, 945 
\bibitem[Evans \& Hawley(1988)]{1988ApJ...332..659E} Evans, C.~R., \& \ Hawley, J.~F.\ 1988, \apj, 332, 659 
\bibitem[Fiedler \& Mouschovias(1993)]{1993ApJ...415..680F} Fiedler, R.~A., \& Mouschovias, T.~C.\ 1993, \apj, 415, 680 
\bibitem[Fromang et al.(2004)]{2004ApJ...616..364F} Fromang, S., Balbus, S.~A., Terquem, C., \& De Villiers, J.-P.\ 2004, \apj, 616, 364 
\bibitem[Fromang et al.(2006)Fromang, Hennebelle, \& Teyssier]{2006A&A...457..371F} Fromang, S., Hennebelle, P., \& Teyssier, R.\ 2006, \aap, 457, 371 
\bibitem[Galli et al.(2006)]{2006ApJ...647..374G} Galli, D., Lizano, S., Shu, F.~H., \& Allen, A.\ 2006, \apj, 647, 374 
\bibitem[Galli \& Shu(1993)]{1993ApJ...417..243G} Galli, D., \& Shu, F.~H.\ 1993, \apj, 417, 243 
\bibitem[Girart et al.(2006)Girart, Rao, \& Marrone]{2006Sci...313..812G} Girart, J.~M., Rao, R., \& Marrone, D.~P.\ 2006, Science, 313, 812 
\bibitem[Goodman et al.(1993)]{1993ApJ...406..528G} Goodman, A.~A., Benson, P.~J., Fuller, G.~A., \& Myers, P.~C.\ 1993, \apj, 406, 528 
\bibitem[Gueth \& Guilloteau(1999)]{1999A&A...343..571G} Gueth, F., \& Guilloteau, S.\ 1999, \aap, 343, 571 
\bibitem[Hawley \& Krolik(2001)]{2001ApJ...548..348H} Hawley, J.~F., \& Krolik, J.~H.\ 2001, \apj, 548, 348 
\bibitem[Heiles \& Troland(2005)]{2005ApJ...624..773H} Heiles, C., \& Troland, T.~H.\ 2005, \apj, 624, 773 
\bibitem[Hennebell \& Fromang(2007)]{2007A&A_HF_sub} Hennebell, P., \& Fromang, S.\ 2007, \aap, submitted
\bibitem[Hennebell \& Teyssier(2007)]{2007A&A_HT_sub} Hennebell, P., \& Teyssier, R.\ 2007, \aap, submitted
\bibitem[Houde(2004)]{2004ApJ...616L.111H} Houde, M.\ 2004, \apjl, 616, L111 
\bibitem[Kenyon \& Hartmann(1995)]{1995ApJS..101..117K} Kenyon, S.~J., \& Hartmann, L.\ 1995, \apjs, 101, 117 
\bibitem[K{\"o}nigl(1991)]{1991ApJ...370L..39K} K{\"o}nigl, A.\ 1991, \apjl, 370, L39 
\bibitem[K{\"o}nigl \& Pudritz(2000)]{2000prpl.conf..759K} K{\"o}nigl, A., \& Pudritz, R.~E.\ 2000, in Protostars and Planets IV, ed. V. Manning, A. Boss, \& S. Russell (Arizona: Univ of Arizona Press), 759 
\bibitem[Krasnopolsky \& K{\"o}nigl(2002)]{2002ApJ...580..987K} Krasnopolsky, R., K{\"o}nigl, A.\ 2002, \apj, 580, 987 
\bibitem[Krumholz et al.(2007)Krumholz, Klein, \& McKee]{2007ApJ...656..959K} Krumholz, M.~R., Klein, R.~I., \& McKee, C.~F.\ 2007, \apj, 656, 959 
\bibitem[Lee et al.(2005)Lee, Ho, \& White]{2005ApJ...619..948L} Lee, C.-F., Ho, P.~T.~P., \& White, S.~M.\ 2005, \apj, 619, 948 
\bibitem[Li \& McKee(1996)]{1996ApJ...464..373L} Li, Z.-Y., \& McKee, C.~F.\ 1996, \apj, 464, 373 
\bibitem[Li \& Shu(1996)]{1996ApJ...472..211L} Li, Z.-Y., \& Shu, F.~H.\ 1996, \apj, 472, 211 
\bibitem[Li \& Shu(1997)]{1997ApJ...475..237L} Li, Z.-Y., \& Shu, F.~H.\ 1997, \apj, 475, 237 
\bibitem[Lizano \& Shu(1989)]{1989ApJ...342..834L} Lizano, S., \& Shu, F.~H.\ 1989, \apj, 342, 834 
\bibitem[Lynden-Bell(2003)]{2003MNRAS.341.1360L} Lynden-Bell, D.\ 2003, \mnras, 341, 1360 
\bibitem[Machida et al.(2007)Machida, Inutsuka, \& Matsumoto]{2007ApJ_MIM_sub} Machida, M.~N., Inutsuka, S.-i., \& Matsumoto, T.\ 2007, \apj, submitted 
\bibitem[Matzner \& McKee(2000)]{2000ApJ...545..364M} Matzner, C.~D., \& McKee, C.~F.\ 2000, \apj, 545, 364 
\bibitem[McKee \& Tan(2002)]{2002Natur.416...59M} McKee, C.~F., \& Tan, J.~C.\ 2002, \nat, 416, 59 
\bibitem[McKee et al.(1993)]{1993prpl.conf..327M} McKee, C.~F., Zweibel, E.~G., Goodman, A.~A., \& Heiles, C.\ 1993, in Protostars and Planets III, ed. E.~H. Levy \& J. Lunine (Tucson:Univ of Arizona Press), 327 
\bibitem[Mestel \& Paris(1979)]{1979MNRAS.187..337M} Mestel, L., \& Paris, R.~B.\ 1979, \mnras, 187, 337 
\bibitem[Mouschovias \& Ciolek(1999)]{1999osps.conf..305M} Mouschovias, T.~C., \& Ciolek, G.~E.\ 1999,in NATO ASIC Proc.~540, The Origin of Stars and Planetary Systems, ed Lada, C.~J. and Kylafis, N.~D. (Norwell, MA: Kluwer Acad. Pub.), 305 
\bibitem[Mouschovias \& Paleologou(1979)]{1979ApJ...230..204M} Mouschovias, T.~C., \& Paleologou, E.~V.\ 1979, \apj, 230, 204 
\bibitem[Myers(1995)]{1995mcsf.conf...47M} Myers, P. C. 1995, in Molecular Clouds and Star Formation, ed. C. Yuan \& J. You (Singapore: World Scientific), 47
\bibitem[Nakano(1989)]{1989MNRAS.241..495N} Nakano, T.\ 1989, \mnras, 241, 495 
\bibitem[Nakano et al.(2002)Nakano, Nishi, \& Umebayashi]{2002ApJ...573..199N} Nakano, T., Nishi, R., \& Umebayashi, T.\ 2002, \apj, 573, 199 
\bibitem[Onishi et al.(2002)]{2002ApJ...575..950O} Onishi, T., Mizuno, A., Kawamura, A., Tachihara, K., \& Fukui, Y.\ 2002, \apj, 575, 950 
\bibitem[Pessah \& Psaltis(2005)]{2005ApJ...628..879P} Pessah, M.~E., \& Psaltis, D.\ 2005, \apj, 628, 879 
\bibitem[Plume et al.(1997)]{1997ApJ...476..730P} Plume, R., Jaffe, D.~T., Evans, N.~J., II, Martin-Pintado, J., \& Gomez-Gonzalez, J.\ 1997, \apj, 476, 730 
\bibitem[Shu(1977)]{1977ApJ...214..488S} Shu, F.~H.\ 1977, \apj, 214, 488 
\bibitem[Shu et al.(2006)]{2006ApJ...647..382S} Shu, F.~H., Galli, D., Lizano, S., \& Cai, M.\ 2006, \apj, 647, 382 
\bibitem[Shu et al.(1987)Shu, Adams, \& Lizano]{1987ARA&A..25...23S} Shu, F.~H., Adams, F.~C., \& Lizano, S.\ 1987, \araa, 25, 23 
\bibitem[Shu et al.(2000)]{2000prpl.conf..789S} Shu, F.~H., Najita, J.~R., Shang, H., \& Li, Z.-Y.\ 2000, in Protostars and Planets IV, ed. V. Manning, A. Boss, \& S. Russell (Arizona: Univ of Arizona Press), 789 
\bibitem[Stone \& Norman(1992a)]{1992ApJS...80..753S} Stone, J.~M., \& Norman, M.~L.\ 1992, \apjs, 80, 753 
\bibitem[Stone \& Norman(1992b)]{1992ApJS...80..791S} Stone, J.~M., \& Norman, M.~L.\ 1992, \apjs, 80, 791 
\bibitem[Stone \& Pringle(2001)]{2001MNRAS.322..461S} Stone, J.~M., \& Pringle, J.~E.\ 2001, \mnras, 322, 461 
\bibitem[Tafalla et al.(1998)]{1998ApJ...504..900T} Tafalla, M., Mardones, D., Myers, P.~C., Caselli, P., Bachiller, R., \& Benson, P.~J.\ 1998, \apj, 504, 900 
\bibitem[Tafalla et al.(2002)]{2002ApJ...569..815T} Tafalla, M., Myers, P.~C., Caselli, P., Walmsley, C.~M., \& Comito, C.\ 2002, \apj, 569, 815 
\bibitem[Tassis \& Mouschovias(2005)]{2005ApJ...618..783T} Tassis, K., \& Mouschovias, T.~C.\ 2005, \apj, 618, 783 
\bibitem[Terebey et al.(1984)Terebey, Shu, \& Cassen]{1984ApJ...286..529T} Terebey, S., Shu, F.~H., \& Cassen, P.\ 1984, \apj, 286, 529 
\bibitem[Tomisaka(1998)]{1998ApJ...502L.163T} Tomisaka, K.\ 1998, \apjl, 502, L163 
\bibitem[Ziegler(2005)]{2005A&A...435..385Z} Ziegler, U.\ 2005, \aap, 435, 385 
\end{thebibliography}
\end{document}